\documentclass[10pt,pre,reprint,amsmath,amssymb,groupedaddress,notitlepage,twocolumn,longbibliography]{revtex4-2}

\makeatletter
\renewcommand*{\p@subsubsection}{}
\makeatother

\usepackage{bm,amsmath,amssymb
}
\usepackage{amsmath}
\usepackage{graphicx,epsfig}
\usepackage{physics}
\usepackage{stmaryrd}
\usepackage{mathtools}
\usepackage{mathtools}
\usepackage{float}
\usepackage{empheq}
\usepackage{epstopdf}
\usepackage{epstopdf}
\usepackage{amssymb}
\usepackage{wrapfig}
\usepackage{picture}
\usepackage{mwe,tikz}
\usepackage[linktoc=all]{hyperref}
\usepackage{color}
\usepackage{transparent}
\usepackage{textcomp}
\usepackage{overpic}
\usepackage{mathrsfs}
\usepackage{natbib}
\makeatletter
\newcommand{\verbatimfont}[1]{\def\verbatim@font{#1}}%
\makeatother

\definecolor{Lblu}{RGB}{193,193,234}
\definecolor{Mblu}{RGB}{131,131, 221}
\definecolor{Dblu}{RGB}{0,0, 255}
\definecolor{Lgr}{RGB}{192,192,192}
\definecolor{Mgr}{RGB}{138,138,138}

\definecolor{OR}{RGB}{244,127,31} 
\definecolor{BR}{RGB}{170,0,0} 
\definecolor{GR}{RGB}{26,211,26} 
\definecolor{VIO}{RGB}{170,0,170} 
\definecolor{PI}{RGB}{237,159,165} 
\definecolor{fgreen}{RGB}{1,68,33}

    \def\XXint#1#2#3{{\setbox0=\hbox{$#1{#2#3}{\int}$}
         \vcenter{\hbox{$#2#3$}}\kern-.5\wd0}}

\hypersetup{
	colorlinks,
	citecolor=blue,
	filecolor=red,
	linkcolor=blue,
	urlcolor=blue,
	hyperfigures
}

\newcommand{\beq}{\begin{equation}}
\newcommand{\eeq}{\end{equation}}

\begin{document}

\title{{Active transport of a passive colloid in a bath of run-and-tumble particles}}

\author{Tanumoy Dhar}
\author{David Saintillan}
\email{dstn@ucsd.edu}
\affiliation{Department of Mechanical and Aerospace Engineering, University of California San Diego,  9500 Gilman Drive, La Jolla, CA 92093, USA}

\begin{abstract}
The dispersion of a passive colloid immersed in a bath of non-interacting and non-Brownian run-and-tumble microswimmers in two dimensions is analyzed using stochastic simulations and an asymptotic theory, both based on a minimal model of swimmer-colloid collisions characterized solely by frictionless steric interactions. We estimate the effective long-time diffusivity $\mathcal{D}$ of the suspended colloid resulting from its interaction with the active bath, and elucidate its dependence on the level of activity (persistence length of swimmer trajectories), the mobility ratio of the colloid to a swimmer, and the number density of swimmers in the bath. We also propose a semi-analytical model for the colloid diffusivity in terms of the variance and correlation time of the net fluctuating active force on the colloid resulting from swimmer collisions. Quantitative agreement is found between numerical simulations and analytical results in the experimentally-relevant regime of low swimmer density, low mobility ratios, and high activity.
\end{abstract}

\maketitle
\section{{Introduction}}

Systems driven out of equilibrium are ubiquitous, and underlying active fluctuations play an important role in governing their behavior. One such example is the ability of active microswimmers to transport colloids \cite{wu2000particle,hernandez2005transport,mino2011enhanced,grober2023unconventional} and perform mechanical work \cite{angelani2009self,di2010bacterial,sokolov2010swimming,kaiser2014transport}. The sole propulsion of individual microswimmers can lead to novel transport characteristics, such as enhanced diffusion \cite{kim2004enhanced,underhill2008diffusion,kurtuldu2011enhancement,jepson2013enhanced}, where the long-time diffusivity of suspended passive particles in bacterial suspensions far exceeds its Brownian counterpart. Seminal experiments by \citet{wu2000particle} first highlighted this phenomenon and uncovered a transition from short-time directed motion (non-Gaussian and super-diffusive) to long-time diffusive (Gaussian) motion.\ A succession of subsequent experiments \cite{kim2004enhanced,chen2007fluctuations,mino2011enhanced,valeriani2011colloids,mino2013induced,jepson2013enhanced, ortlieb2019statistics, maggi2017memory, lagarde2020colloidal} further probed this effect, and for instance examined the role of the concentration of bacteria and the effect of the size ratio of the colloid to the bacteria. The type of microswimmer used was also shown to affect the dynamics of the colloids suspended in a bath: for example, certain algae (\textit{C. reinhardtii}) drive oscillatory flows \cite{guasto2010oscillatory,drescher2010direct} resulting in loop-like colloid trajectories \cite{leptos2009dynamics}, an effect that can be rationalized using far-field hydrodynamics of this swimming organism \cite{dunkel2010swimmer,klindt2015flagellar,jeanneret2016entrainment}. An extensive amount of theoretical \cite{mino2011enhanced,jepson2013enhanced,kasyap2014hydrodynamic,morozov2014enhanced, burkholder2017tracer,maes2020fluctuating, solon2022einstein,kanazawa2020loopy} and numerical works \cite{saintillan2008instabilities,hernandez2005transport,underhill2008diffusion,lin2011stirring,morozov2014enhanced} were performed over the last two decades, and several models were proposed to elucidate the respective roles of steric \citep{burkholder2017tracer,lagarde2020colloidal} vs hydrodynamic \citep{underhill2008diffusion,lin2011stirring,morozov2014enhanced,thiffeault2015distribution} effects in driving colloidal diffusion. 
Recent experiments have shown that bacteria \cite{brown2016swimming,grober2023unconventional} and other self-propelled particles \cite{kummel2015formation,brown2016swimming,ramananarivo2019activity} can undergo direct collisions with colloidal clusters, substantiating the significance of steric interactions relative to hydrodynamic interactions. However, modeling endeavors focused on predicting colloid dispersion in active suspensions have encountered various limitations, primarily attributable to the challenges involved in quantifying the scattering dynamics between the active particle and the colloid. Recently, \citet{lagarde2020colloidal} put forth a collision model that treats bacteria-colloid interactions as purely repulsive. The present work is inspired by the scattering models of \citet{jakuszeit2019diffusion} and \citet{saintillan2023dispersion}, used to study the dispersion of an active particle in porous media. Regardless of the nature of interactions, the anomalous short-time dynamics exhibited by colloids suspended in bacterial baths are always found to map to Gaussian statistics at long times. The quantitative understanding of this microscopic enhancement of the long-time diffusivity of passive colloids in active baths remains, however, incomplete, even in the seemingly simple case of steric interactions. Previous works \cite{mino2011enhanced,mino2013induced,jepson2013enhanced,morozov2014enhanced} have provided phenomenological arguments for such diffusion. Yet, an analytical model relating the diffusion coefficient to system parameters based on first principles is still lacking for baths of run-and-tumble particles. 

To address this gap, we propose a minimal model incorporating swimmer-colloid collisions in two dimensions (where the majority of the experiments \cite{wu2000particle,valeriani2011colloids,mino2011enhanced,patteson2016particle, lagarde2020colloidal,grober2023unconventional} are performed) to effectively capture the effect of system parameter variations on the dispersion. The proposed model relies on a collision-resolution algorithm that solves for constraint forces arising from swimmer-colloid contacts at every instant in time. We employ stochastic simulations to estimate the effective diffusivity $\mathcal{D}$ of the suspended colloid, highlighting its dependence on the level of activity or P\'eclet number $Pe$, mobility ratio $\mu$ of the colloid to a swimmer, and number density $\rho$ of swimmers in the active bath. These system parameters are defined more precisely later. We also posit a semi-analytical model for the colloid diffusivity in terms
of the variance and correlation time of the fluctuating active force resulting from swimmer collisions. Our theoretical predictions agree well with particle simulations in the limit of low swimmer density, low mobility ratio and high P\'eclet number, and provide a simple analytical framework for the description of long-time colloidal dispersion in baths of run-and-tumble microswimmers.

The paper is organized as follows. In Section~\ref{section2}, we present the swimmer-colloid collision model, the dimensionless governing equations, and the computational framework. We present results from stochastic simulations in Section~\ref{section4}, where we highlight the dependence of the long-time colloid diffusivity on the system parameters. In Section~\ref{section5}, we propose an asymptotic theory valid for low mobility ratios $(\mu \longrightarrow 0)$. This theory relates the diffusivity to the variance and correlation time of the fluctuating force on a fixed colloid subject to collisions with run-and-tumble microswimmers, and involves the determination of the mean number of swimmers in contact with the colloid. The proposed semi-analytical theory is shown to match well with the numerical simulation results in the limit of large colloids and long swimmer persistence lengths. We conclude in Section~\ref{conclusions}.

\section{Problem formulation}\label{section2}
\subsection{Swimmer-colloid collision model}\label{section2a}

\begin{figure}[t]
\centering
\includegraphics[width=0.45\textwidth]{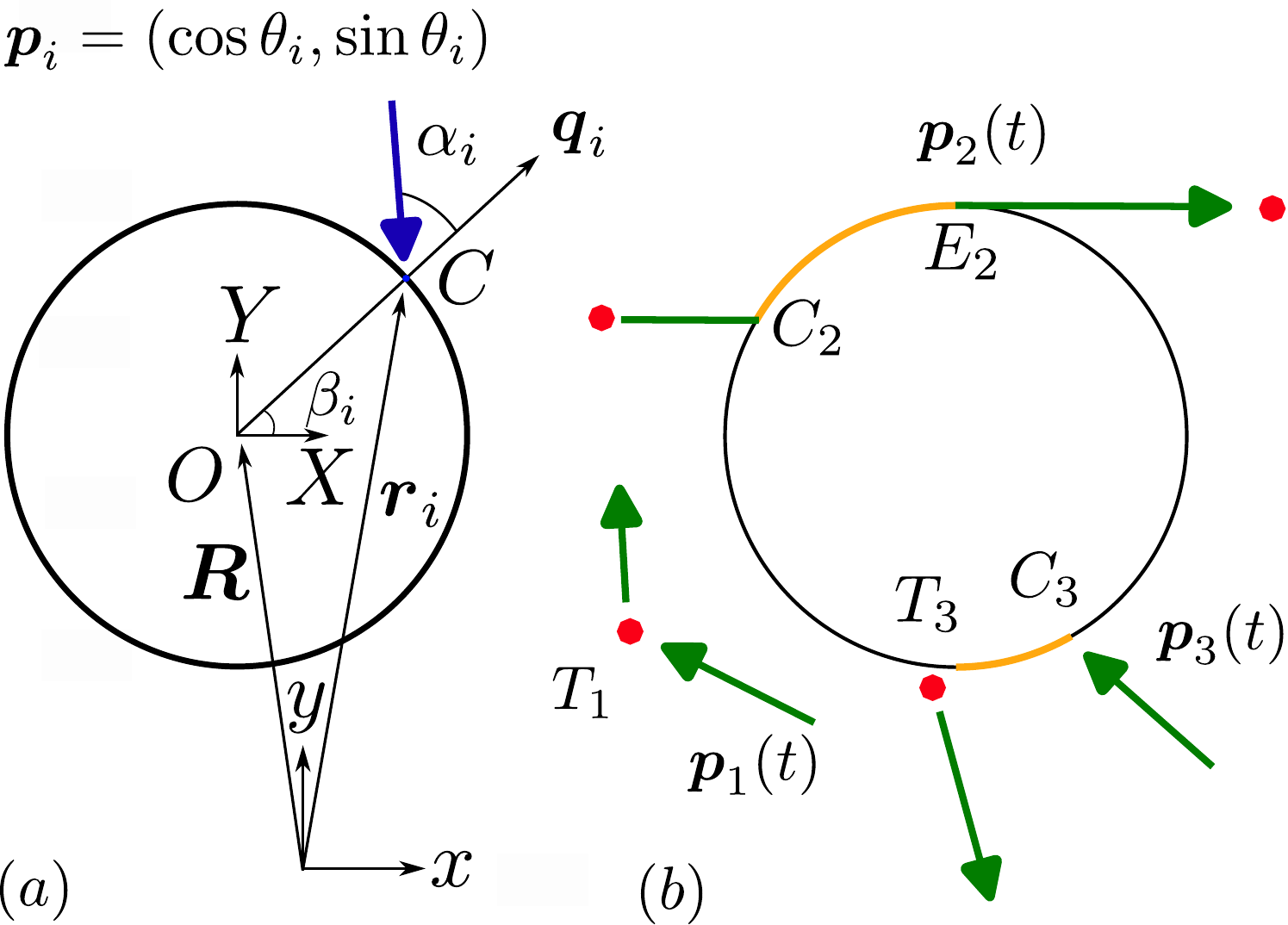}
\caption{Problem definition: (a) Schematic diagram of a swimmer in contact with the colloid and associated nomenclature (see main text for details); (b)~Representative swimmer dynamics, where sections of a run performed in the bulk or on the colloid surface are shown in green and yellow, respectively, and where tumbles are depicted as red dots. As a swimmer slides on the colloid surface, it can either escape by swimming away tangentially (swimmer 2), or finish its run and perform its next tumble on the surface (swimmer 3).}
\label{schematic}
\end{figure}

We analyze the dispersion of a passive colloid suspended in a dilute collection of non-interacting self-propelled particles in two dimensions using a minimal model for swimmer-colloid collisions. 
Swimmers ($i=1,\dots,N$), with positions $\boldsymbol{r}_i$ and orientations $\boldsymbol{p}_i$ (where $|\boldsymbol{p}_i|=1$), are modeled as point particles and perform a run-and-tumble motion \cite{berg1977physics,berg2018random} away from the colloid: straight swimming runs of duration $\tau$ with constant velocity $u$ along the unit director $\boldsymbol{p}_i$ are interrupted with instantaneous reorientation events or tumbles (see Fig.~\ref{schematic}b where the swimmer with orientation $\boldsymbol{p}_{1}(t)$ performs a tumble at a point denoted by $T_{1}$). The time $\tau$ between consecutive tumbles is an exponentially distributed random variate with mean $\bar{\tau}$ and cumulative probability distribution function $\Sigma(\tau) = 1 - \exp(-\tau/\bar{\tau})$ for $\tau \geq 0$. The choice of an exponential distribution to model the tumbling rate of the \textit{Escherichia coli} bacterium \citep{block1983adaptation,alon1998response,saragosti2011directional} is an assumption and has been widely used in previous theoretical studies \cite{soto2014run,elgeti2015run}.  In the present work, we assume that the swimming velocity $u$, orientation $\boldsymbol{p}_i$ and run time $\tau$ of the swimmer remain independent of its spatial proximity to the colloid, which serves as a convenient approximation for computational simulations. In free space, far from the colloid, a swimmer performs unimpeded run-and-tumble motion. During a run away from the colloid, the swimmer dynamics simply follows:
\begin{equation}\label{eq01}
\dot{\boldsymbol{r}}_{i} = u\boldsymbol{p}_{i}.
\end{equation}

A steric collision occurs whenever a swimmer comes in contact with the surface of the colloid, assumed to be of circular shape with radius $A$. The geometry of a collision is illustrated in the schematic of Fig.~\ref{schematic}a. We choose a Cartesian $(x,y)$ coordinate system that is fixed in the active bath, where $\boldsymbol{r}_{i}$ and $\boldsymbol{R}$ are the instantaneous swimmer and colloid positions, respectively. The current orientation of the swimmer is $\boldsymbol{p}_{i} = (\cos\theta_{i},\sin \theta_{i})$, where the angle $\theta_{i}$ during a particular run is uniformly distributed in the interval $[0,2\pi)$.
The wall-normal unit vector $\boldsymbol{q}_{i}$ at the point of contact $C$ is directed from the colloid into the active bath. The angle $\alpha_{i}\in[-\pi,\pi]$ in Fig.~\ref{schematic}a captures the projection of $\boldsymbol{p}_{i}$ on $\boldsymbol{q}_{i}$ and is henceforth called contact angle. Note that $\alpha_i$ changes with time over the course of a run as the swimmer slides over the colloid surface. Symmetry with respect to the direction $\boldsymbol{q}_i$ dictates that the probability density function of $\alpha_i$ be an even function. Finally, the angle $\beta_{i}\in[0,2\pi)$ captures the angular position of the swimmer on the colloid surface with respect to the $(X,Y)$ coordinate system.


We model swimmer-colloid interactions as steric frictionless collisions: 
When a swimmer comes in contact with the colloid, its relative velocity with respect to the colloid vanishes in the normal direction, effectively establishing a no-penetration boundary condition. By assumption, the swimmer orientation remains unchanged for the duration of the current run, and it simply undergoes sliding motion along the colloid surface until it either swims or tumbles away. If the contact angle reaches $\pm \pi/2$ before the end of the run, the swimmer escapes from the colloid surface and completes its run in a straight line in the bulk (see Fig.~\ref{schematic}b where swimmer 2 comes in contact with the colloid at point $C_{2}$ and escapes at point $E_{2}$ where $\boldsymbol{p}_{2}$ becomes tangent to the surface). If, on the other hand,  the swimmer concludes its run on the surface of the colloid, it will perform its next tumble on the surface. If its new orientation points into the bulk, it will immediately leave the colloid surface (see swimmer 3 in Fig.~\ref{schematic}b). If its new orientation points toward the colloid, it will start the next run sliding on the colloid surface, until it either escapes or tumbles again.


During the interaction between the swimmer and the colloid, both entities apply contact forces $\pm F_{i}\boldsymbol{q}_i$ of equal magnitude but in opposite directions. As the swimmer slides on the surface of the colloid, the contact force magnitude changes with the swimmer's evolving position and orientation.\  Tangential resistance to sliding is not considered, resulting in a contact force directed solely along the normal direction. The swimmer's velocity is modified as a result of the collision:
\begin{equation}\label{eq04}
\dot{\boldsymbol{r}}_{i} = u\boldsymbol{p}_{i} + m F_{i}\boldsymbol{q}_{i},
\end{equation}
where $m$ is the mobility of the swimmer, assumed to be isotropic for simplicity. The dynamics of the colloid is driven by the net contact force resulting from the collisions with any swimmers currently on its surface and follows
\begin{equation}\label{eq05}
\dot{\boldsymbol{R}} = -M\sum_{i=1}^{c}F_{i}\boldsymbol{q}_{i},
\end{equation}
where $M$ is the mobility of the colloid. Here, $c(t)$ is the number of swimmers in contact with the colloid at any given instant in time. As we explain in detail in Section~\ref{sec:numerics}, the contact forces $F_{i}$ are determined numerically to satisfy the no-penetration condition:
\begin{equation}\label{nopenetrationbc}
(\dot{\boldsymbol{r}}_{i} - \dot{\boldsymbol{R}})\cdot \boldsymbol{q}_{i} = 0.
\end{equation}
We also introduce $\zeta = 1/m$ and $Z = 1/M$ as the friction coefficients of a swimmer and of the colloid, respectively. Throughout the remainder of the present work, we choose the convention of employing uppercase letters to represent physical variables associated with the passive colloid, and lowercase letters for those associated with a swimmer.

We underscore that the minimal model presented herein provides a simple framework for understanding colloidal dispersion in active baths, but also relies on strong simplifying assumptions: in the context of experiments, short-ranged hydrodynamic effects (lubrication \cite{drescher2011fluid,stenhammar2017role,yoshinaga2018hydrodynamic} and alignment torques \cite{berke2008hydrodynamic,lauga2006swimming}) experienced by the swimmers can lead to scattering at angles that are often non-tangent to the colloid surface \cite{spagnolie2015geometric}.\ Experiments demonstrate that the tumbling rate $1/\tau$ is reduced for certain swimmers such as \textit{E. coli} near boundaries, which effectively acts towards trapping the swimmer at the walls \cite{molaei2014failed,perez2019bacteria}.\
Finally, the present model does not capture the convex trajectory of the swimmer around the colloid predicted by models that account for swimmer-colloid hydrodynamic interactions \cite{spagnolie2015geometric,sipos2015hydrodynamic}, but is supported by the recent experimental observations of \citet{lagarde2020colloidal}. 

\subsection{Dimensionless governing equations} 
The governing equations are non-dimensionalized using length scale $A$ and time scale $A/u$.
Contact forces are made dimensionless using the swim force \cite{takatori2014swim,takagi2014hydrodynamic,omar2020microscopic} $F^{\text{swim}} = \zeta u$, which is the force one needs to apply to a swimmer to fix its position in space. In free space, away from the colloid, the dynamics of a swimmer during its run simply follows
\begin{equation}\label{eq01nd}
\boldsymbol{\dot{r}}_{i} = \boldsymbol{p}_{i}.
\end{equation}
At the colloid surface, it becomes:
\begin{equation}\label{eq02nd}
\dot{\boldsymbol{r}_{i}} = \boldsymbol{p}_{i} + F_{i}\boldsymbol{q}_{i}.
\end{equation}
The evolution equation for the position of the colloid reads:
\begin{equation}\label{eq04a}
\dot{\boldsymbol{R}} = -\mu\sum_{i=1}^{c}F_{i}\boldsymbol{q}_{i},
\end{equation}
where the mobility ratio $\mu$ is defined as
\begin{equation}\label{eq072}
\mu = \frac{M}{m} = \frac{\zeta}{Z}.
\end{equation}

In addition to the mobility ratio, two dimensionless parameters govern the problem. The P\'eclet number, defined as $Pe = u\bar{\tau}/A$, compares the mean run time $\bar{\tau}$ to the time $A/u$ for a swimmer to travel a distance of one colloid radius, and is a dimensionless measure of the persistence of swimmer trajectories. Note that $Pe$ does not appear explicitly in the governing equations, but defines the mean of the exponential distribution used to model the swimmer run times. Finally, the number density $\rho$ of swimmers is defined as the ratio of the total number of swimmers $\nu$ in the square domain of length $L$ to the area $L^{2}-\pi A^{2}$ available to the swimmers. In dimensionless variables,
\begin{equation}
\rho = \frac{\nu A^{2}}{L^{2}-\pi A^{2}}. \label{eq:rho}
\end{equation}

\subsection{Numerical method \label{sec:numerics}}
\label{section3}
We implement a contact-force resolution-based computational algorithm for simulating the dynamics of a single passive mobile colloid in a suspension of non-interacting run-and-tumble particles in two dimensions. The simulations are performed inside a periodic square domain of linear dimension $L$ chosen to be much larger than the colloid size and typical run length ($L\gg \mathrm{max}\{Pe,1\}$ in dimensionless variables), which $\nu$ swimmers are distributed at random. The values of $L$ and $\nu$ are chosen to achieve the desired number density $\rho$ as defined in eqn~(\ref{eq:rho}). We discretize time with a fixed time step $\Delta t \ll \mathrm{min}\{Pe,1\}$ and use time-marching to advance the positions of the swimmers and colloid. 

Far from the colloid, a swimmer trajectory simply evolves as $\boldsymbol{r}_{i}(t+\Delta t) = \boldsymbol{r}_{i}(t) + \boldsymbol{p}_{i}\Delta t$ over the course of one time step. During any given time step, a number $c(t)$ of swimmers will collide with the colloid. The positions of these swimmers are updated as
\begin{equation}
\boldsymbol{r}_{i}(t+\Delta t) = \boldsymbol{r}_{i}(t) +  [\boldsymbol{p}_{i} + F_{i}(t)\boldsymbol{q}_{i}] \Delta t, \label{eq:swimmerEuler}
\end{equation}
where the normal contact force $F_i$, taken to be constant over one time step, must be determined. During that same time step, the position $\boldsymbol{R}$ of the colloid is evolved as   \vspace{-0.2cm}
\begin{equation}
\boldsymbol{R}(t+\Delta t) = \boldsymbol{R}(t)  -\mu\sum_{i=1}^{c(t)}F_{i}(t)\boldsymbol{q}_{i} \Delta t,  \label{eq:colloidEuler}
\end{equation}
which couples all the contact forces. 
To solve for the magnitude of these forces, we apply the no-penetration condition of eqn~(\ref{nopenetrationbc}) at the location of each contact. This leads to the following set of coupled linear equations:
\begin{equation}
(\boldsymbol{\mathscr{I}}+\mu \boldsymbol{\mathscr{A}})\cdot \boldsymbol{\mathscr{F}} = -\boldsymbol{\mathscr{B}},  \label{eq:Fsystem}
\end{equation}
where, $\boldsymbol{\mathscr{F}}= \{F_{1}, F_{2}...F_{c}\}$ is a vector of length $c(t)$ containing the unknown contact forces, and $\boldsymbol{\mathscr{I}}$ is the identity matrix. The matrix $\boldsymbol{\mathscr{A}}$ and the vector $\boldsymbol{\mathscr{B}}$ are geometric quantities that depend on the current configuration of the swimmers on the colloid surface:
\begin{align}
\begin{split}
\boldsymbol{\mathscr{A}} = \begin{Bmatrix}
\boldsymbol{q}_{1}\cdot\boldsymbol{q}_{1} & \boldsymbol{q}_{1}\cdot\boldsymbol{q}_{2} & \boldsymbol{q}_{1}\cdot\boldsymbol{q}_{3}& ... & \boldsymbol{q}_{1}\cdot\boldsymbol{q}_{c}\\
\boldsymbol{q}_{2}\cdot\boldsymbol{q}_{1} & \boldsymbol{q}_{2}\cdot\boldsymbol{q}_{2} & \boldsymbol{q}_{2}\cdot\boldsymbol{q}_{3} & ... & \boldsymbol{q}_{2}\cdot\boldsymbol{q}_{c}\\
\boldsymbol{q}_{3}\cdot\boldsymbol{q}_{1} & \boldsymbol{q}_{3}\cdot\boldsymbol{q}_{2} & \boldsymbol{q}_{3}\cdot\boldsymbol{q}_{3} & ... & \boldsymbol{q}_{3}\cdot\boldsymbol{q}_{c}\\
\vdots & \vdots & \vdots & \ddots & \vdots \\
\boldsymbol{q}_{c}\cdot\boldsymbol{q}_{1} & \boldsymbol{q}_{c}\cdot\boldsymbol{q}_{2} & \boldsymbol{q}_{c}\cdot\boldsymbol{q}_{3} & ... & \boldsymbol{q}_{c}\cdot\boldsymbol{q}_{c}
\end{Bmatrix},  \\
\boldsymbol{\mathscr{B}} = \begin{Bmatrix}
\boldsymbol{p}_{1}\cdot\boldsymbol{q}_{1} \\
\boldsymbol{p}_{2}\cdot\boldsymbol{q}_{2}\\
\boldsymbol{p}_{3}\cdot\boldsymbol{q}_{3}\\
\vdots\\
\boldsymbol{p}_{c}\cdot\boldsymbol{q}_{c}
\end{Bmatrix}.\qquad\qquad\qquad
\end{split}
\end{align}
The system of equations~(\ref{eq:Fsystem}) is constructed at every time step and inverted numerically to obtain the contact forces. The swimmer and colloid positions are subsequently updated using eqn~(\ref{eq:swimmerEuler}) and eqn~(\ref{eq:colloidEuler}), respectively. Note that the present formulation can still result in occasional swimmer-colloid overlaps $(|\boldsymbol{r}_{i} - \boldsymbol{R}| < 1)$ due to the time discretization. As a result, we check for overlaps at the end of each time step and, if necessary,  implement the additional correction for the swimmer position based on the updated colloid location:
\begin{equation}
\boldsymbol{r}_{i}(t+\Delta t) = \boldsymbol{R}(t+\Delta t) - \boldsymbol{q}_{i}.
\end{equation}

\section{Numerical results and discussion}  \label{section4}

\begin{figure}[t]
\begin{center}
\includegraphics[width=0.33\textwidth]{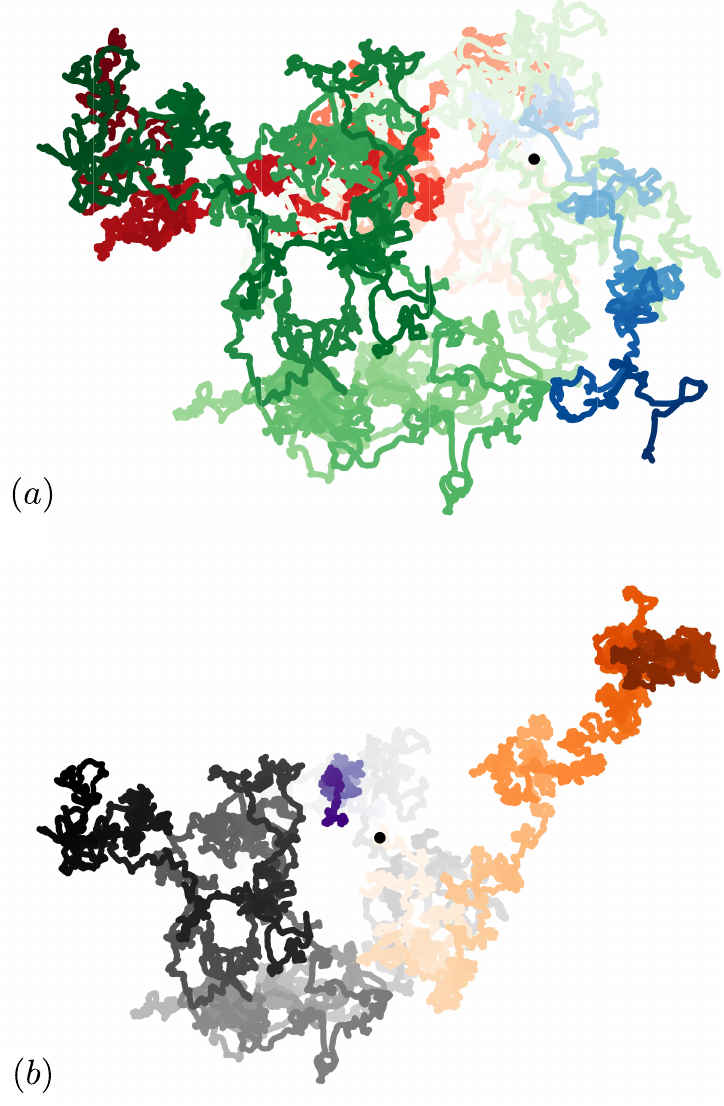}
\end{center}\vspace{-0.3cm}
\caption{Simulated trajectories of the passive colloid for different parameter values. The three distinct color gradients are for different: (a) number densities of swimmers: $\rho = 0.08$ (blue), $0.8$ (green) and $8$ (red) (with $Pe = 5.0$ and $\mu = 0.5$); (b)  P\'eclet numbers: $Pe = 0.05$ (purple), $0.5$ (orange) and $5$ (grey) (with $\rho = 0.8$ and $\mu =0.5$). The increasing intensity of color captures increasing time. All the trajectories are for $t = 0$ to $5000$. The black dot shows the initial colloid location. See ESI for a video showing a typical simulation.} \vspace{-0.1cm}
\label{sample_trajectories}
\end{figure}
\begin{figure}[b!]
\begin{center}\vspace{-0.1cm}
\includegraphics[width=0.48\textwidth]{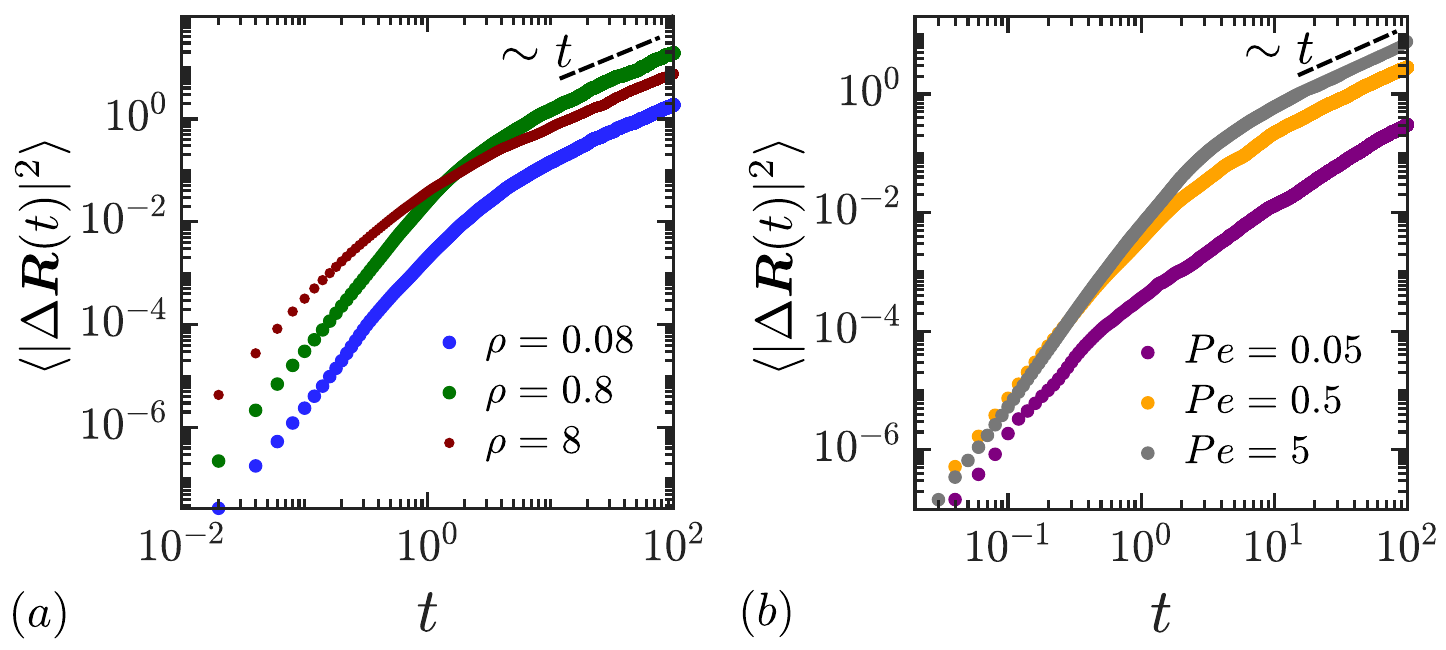} \vspace{-0.6cm}
\caption{Variation of the mean squared displacement with time, for different: (a) number densities of swimmers: $\rho = 0.08$ (blue), $0.8$ (green) and $8$ (red) (with $Pe = 5.0$ and $\mu = 0.5$); (b)  P\'eclet numbers: $Pe = 0.05$ (purple), $0.5$ (orange) and $5$ (grey) (with $\rho = 0.8$ and $\mu =0.5$). The dashed black lines show the linear diffusive scaling at long times.} \vspace{-0.4cm}
\label{msdvst}
\end{center}
\end{figure}

In this section, we present results from numerical simulations for a wide range of system parameters. Figure \ref{sample_trajectories} shows several representative colloid trajectories (for $t=0$ to $5000$) at different densities $\rho$ and different P\'eclet numbers $Pe$ in panels a and b, respectively (see figure caption for details, and also see ESI for a video of a typical simulation). The trajectories, along which increasing color intensity is used to show the passage of time, are correlated random walks, albeit with a short correlation time as we discuss further below. Visual inspection suggests that dispersion gets stronger with either increasing $\rho$ (compare blue and green curves in a) or $Pe$ (compare purple and orange curves in b), trends that we quantify and explain in the following discussion.


\begin{figure*}[t]
\centering
\includegraphics[width=1.0\textwidth]{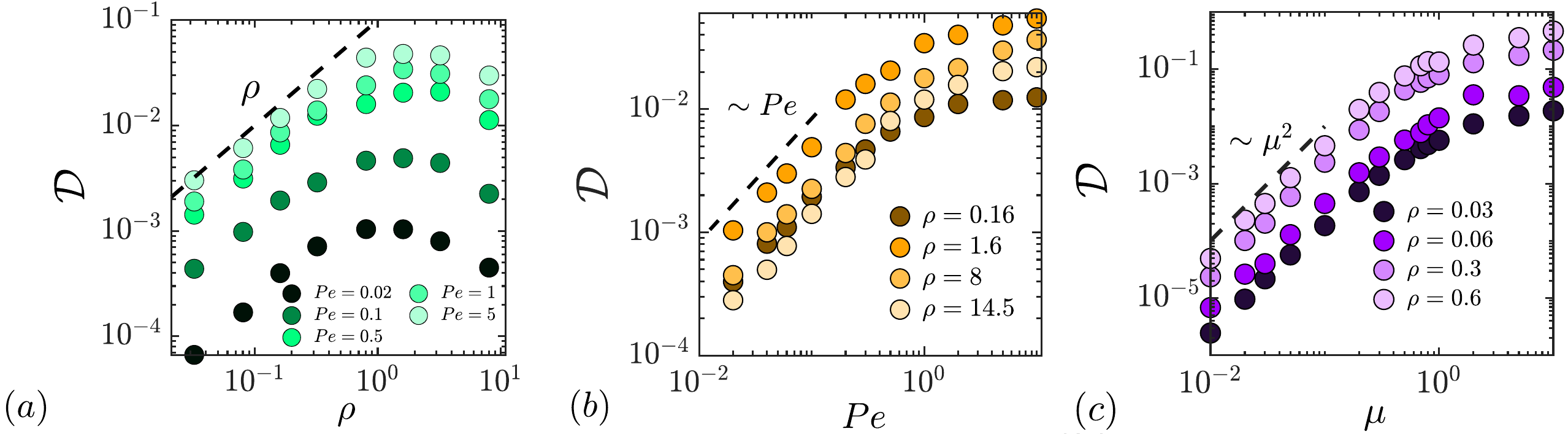}
\caption{Variation of the diffusivity $\mathcal{D}$ of the colloid with:\ (a) number density $\rho$ of swimmers for different values of $Pe$;\ (b) P\'eclet number $Pe$, for the different values of $\rho$;\  (c) mobility ratio $\mu$ of the colloid to the swimmer. The mobility ratio is kept constant at $\mu = 0.5$ in panels a and b.\ The dashed lines in each panel highlight asymptotic scalings at low values of $\rho$, $Pe$ and~$\mu$.}
\label{Fig_diffusivity_variations}
\end{figure*}

Figure \ref{msdvst} shows the time evolution of the mean squared displacement of the colloid, defined as $\langle |\Delta\boldsymbol{R}(t)|^2 \rangle= \langle |\boldsymbol{R}(t)-\boldsymbol{R}(0)|^2 \rangle$, where $\langle \cdot \rangle$ denotes the ensemble average over multiple realizations of single colloid trajectories. The system parameters are the same as in Fig.~\ref{sample_trajectories}. At short time scales ($t \sim 10^{-2}-10^{-1}$), the mean squared displacement grows as $t^{\gamma}$ with $\gamma \approx 2$. At later times, a transition to diffusive behavior (with $\gamma \approx1$) is observed, consistent with previous experimental studies. 
 The crossover time from the ballistic to the diffusive regime varies for different system parameters as is evident in Fig.~\ref{msdvst}a,b. We quantify the long-time dispersion of the colloid in terms of its diffusivity $\mathcal{D}$ defined as:
\begin{equation}
\mathcal{D} = \lim_{t\rightarrow\infty}\frac{1}{4}\frac{\mathrm{d}}{\mathrm{d} t}\langle |\Delta\boldsymbol{R}(t)|^2 \rangle.  \label{eq:Ddef}
\end{equation}
The 
diffusivity was numerically determined for a wide range of number densities $\rho$, P\'eclet numbers $Pe$, and mobility ratios $\mu$, and the results are summarized in Fig.~\ref{Fig_diffusivity_variations}.

The variation of $\mathcal{D}$ with number density $\rho$ is illustrated in Fig.~\ref{Fig_diffusivity_variations}a  for different values of $Pe$ at $\mu=0.5$. At low to moderate swimmer densities, increasing $\rho$ leads to a linear increase of $\mathcal{D}$, which can be rationalized by a concomitant increase in the number of swimmer-colloid collisions as we analyze further below. This linear dependence on $\rho$ is in agreement with previous experimental studies \cite{mino2011enhanced, mino2013induced,lagarde2020colloidal,grober2023unconventional}. Upon further increasing the density of the active bath, the diffusivity $\mathcal{D}$ is found to reach a peak and then slightly decrease with $\rho$ at high densities. A similar departure of the diffusivity from linear growth with increasing swimmer density was previously reported by \citet{lagarde2020colloidal}, in both experiments and numerical simulations. The explanation provided by \citet{lagarde2020colloidal} is based on the hypothesis that multiple bacteria simultaneously interact with the colloid at high densities, resulting in a trend that breaks the linear growth of $\mathcal{D}$. This explanation is consistent with our simulations: the departure from the linear increase in Fig.~\ref{Fig_diffusivity_variations}a occurs for $\rho\gtrsim 1$, which coincides approximately with the density at which the mean number of swimmers on the surface of the colloid at any given time exceeds 1 (see Fig.~\ref{fig_mean_number_swimmers} below). 
The slight decrease with $\rho$ observed at high densities in Fig.~\ref{Fig_diffusivity_variations}a can be understood by the same effect: when many swimmers interact with the colloid at the same time, partial cancellation of the forces they exert on the colloid occurs, resulting in a lower diffusivity; this will be confirmed in Fig.~\ref{force_variance} when we discuss the dependence of the net force variance on $\rho$. 
To the best of our knowledge, this decrease in $\mathcal{D}$ with respect to density at large $\rho$ has not been experimentally observed in bacterial suspensions. Indeed, the regime where it occurs in our model may be challenging to realize experimentally, as it requires moderate to high mobility ratios (i.e., small colloids) as well as many swimmers interacting with the colloid at any given time. In a system with finite-sized swimmers, steric hindrance  would likely impose an upper bound on the number of active particles capable of simultaneously colliding with the colloid surface. Furthermore, bacterial suspensions at such high densities may also exhibit collective motion, which would further act to enhance $\mathcal{D}$.

\begin{figure*}[t]
\centering
\includegraphics[width=0.98\textwidth]{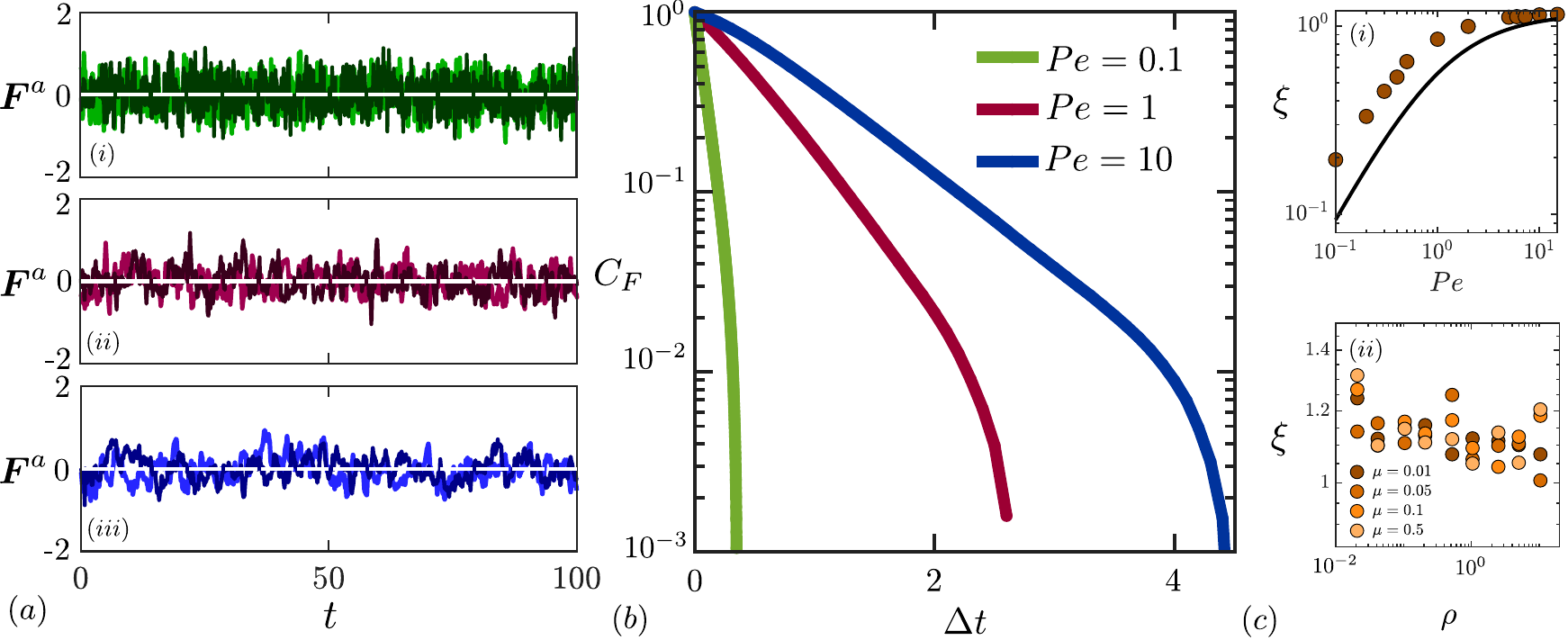}
\caption{(a) Variations of $x$ and $y$ components of the net active force on the colloid with time for P\'eclet numbers $Pe = 0.1$ ($i$), $1$ ($ii$) and $10$ ($iii$), for constant density of swimmers $\rho = 8.04$ and mobility ratio $\mu = 0.5$. (b) Time autocorrelation $C_{F}(\Delta t)$ of the net active force, defined in eqn~(\ref{time_constant_force_autocorr}), for the same conditions as in (a). (c) Time constant $\xi$ of the force autocorrelation function, obtained by exponential fit, as a function of ($i$) P\'eclet number (for $\rho = 1$ and $\mu=0.01$), and $(ii)$ swimmer density (for $Pe=8$ and different mobility ratios). In c($i$), the solid black curve shows the lower-bound estimate derived in eqn~(\ref{eq:xiestimate}). }
\label{force_autocorrelation_time_constant}
\end{figure*}

The dependence of diffusivity on P\'eclet number is shown in Fig.~\ref{Fig_diffusivity_variations}b for different swimmer densities. At lower values of $Pe$ ($Pe \lesssim 1)$, $\mathcal{D}$ increases linearly with $Pe$, since persistent swimmers push on the colloid surface for a longer time before they escape or tumble into the bulk as the P\'eclet number is increased. For sufficiently large P\'eclet number ($Pe \gtrsim 1$), the diffusivity saturates. This can be attributed to the fact that at large values of $Pe$ most collisions result in an escape, and therefore the amount of time that a swimmer interacts with the colloid becomes independent of $Pe$; this trend will be confirmed in Fig.~\ref{force_autocorrelation_time_constant}(c) when we discuss the autocorrelation time for the net force on the colloid. Finally, panel c of Fig.~\ref{Fig_diffusivity_variations} shows the dependence of $\mathcal{D}$ on the mobility ratio $\mu$. In the limit of large colloids ($\mu \lesssim 1$), the dependence is quadratic as $\mathcal{D} \sim \mu^2$, but $\mathcal{D}$ is found to eventually saturate in the limit of small colloids ($\mu \gtrsim 1$). Next, we rationalize some of these trends by deriving a semi-analytical model for the diffusivity in the limits of low swimmer density, high P\'eclet number and low mobility ratio. 
 

\section{Semi-analytical description}
\label{section5} 

In this section, we present an analytical model for the diffusivity of a passive colloid suspended in the active bath. Following \citet{hinch1975application},  we start from the equations of motion for the suspended colloid and show that the effective dispersion coefficient can be related to the time autocorrelation function of the net active force it experiences due to collisions with the active bath. We then show how that autocorrelation function can be estimated analytically in the asymptotic limit of low mobility ratios $(\mu \longrightarrow 0)$, in which we obtain a closed-form expression for the diffusivity. We then compare our theory to results from simulations and find excellent agreement in the limit of validity of the model. 
\subsection{{Diffusivity calculation}}

The run-and-tumble random walk of the active microswimmers translates to fluctuations in the motion of the colloid. 
The starting point for our diffusivity calculation in the absence of Brownian fluctuations is the equation of motion (\ref{eq05}) of the passive colloid, written as 
\begin{equation}
\dot{\boldsymbol{R}}=\mu\boldsymbol{F}^a(t),   \label{eq:colloidvelocity}
\end{equation}
where $\boldsymbol{F}^a(t)$ is the net instantaneous fluctuating active force resulting from collisions with the swimmers:
\begin{equation}
    \boldsymbol{F}^a(t) = -\sum_{i=1}^{c(t)}{F}_i^c(t)\boldsymbol{q}_i(t). \label{eq:activeforce}
\end{equation}
As explained in Sec.~\ref{section2a}, the net force on the colloid results from contact forces with the swimmers on its surface, and thus depends on 
the number $c(t)$ of swimmers in contact,  as well as their positions and orientations on the colloid surface, which determine the magnitude of the contact forces $F_i^c(t)$ and their directions $\boldsymbol{q}_i(t)$. 
Integrating eqn~(\ref{eq:colloidvelocity}) and choosing the origin such that $\boldsymbol{R}(0)=\mathbf{0}$, the displacement of the passive colloid is found to be,
\begin{equation}\label{msd_form1}
\Delta\boldsymbol{R}(t) = \boldsymbol{R}(t)-\boldsymbol{R}(0) = \int_{0}^{t} \mu \boldsymbol{F}^a (t')\, \mathrm{d}t'.
\end{equation}
We use this expression to calculate the mean square displacement as
\begin{equation}
\langle |\Delta\boldsymbol{R}(t)|^2 \rangle = 
\mu^{2}\int_{0}^{t}\int_{0}^{t}\langle\boldsymbol{F}^a(t')\cdot \boldsymbol{F}^a(t'')\rangle\, \mathrm{d} t' \mathrm{d} t'',   \label{eq:msd}
\end{equation}
where the angle brackets $\langle \cdot\rangle$ denote the ensemble average. Eqn~(\ref{eq:msd}) is the well-known Green-Kubo relation \cite{kubo2012statistical}.

We gain intuition on the force autocorrelation function by plotting data from simulations in Fig.~\ref{force_autocorrelation_time_constant}a, showing the $x$ and $y$ components of the active force as functions of time for different P\'eclet numbers. As expected, the net active force fluctuates around zero, $\langle \boldsymbol{F}^a(t) \rangle =  \boldsymbol{0}$. As the P\'eclet number increases, time fluctuations become slower, indicating an increase in the autocorrelation time scale. This is quantified in Fig.~\ref{force_autocorrelation_time_constant}b, where we plot the autocorrelation function $C_{F} (\Delta t)$ defined as
\begin{equation}\label{time_constant_force_autocorr}
C_{F}(\Delta t) = \frac{ \langle \boldsymbol{F}^a(t)\cdot \boldsymbol{F}^a(t+\Delta t)\rangle}{\langle |\boldsymbol{F}^a|^{2}\rangle}, 
\end{equation}
where the denominator $\langle |\boldsymbol{F}^a|^{2}\rangle$ is the active force variance. Initially, the autocorrelation function decays exponentially with time as $C_{F} (\Delta t)=\exp(-\Delta t/\xi)$, followed by a sharp drop at a given value of $\Delta t$ that depends on $Pe$. 
Leveraging this exponential decay, we can rewrite the mean-squared displacement of eqn~(\ref{eq:msd}) as
\begin{align}
\begin{split}
\langle |\Delta\boldsymbol{R}(t)|^2 \rangle & =
2\mu^{2}\int_{0}^{t}(t-t')\langle |\boldsymbol{F}^a|^{2} \rangle \exp(-t'/\xi)\, \mathrm{d} t', \\
& = 2(\xi \mu)^{2}\langle |\boldsymbol{F}^a|^{2} \rangle \left[\exp(-t/\xi)+(t/\xi)-1\right].
\end{split}
\end{align}
Eqn~(\ref{eq:Ddef}) then provides the long-time diffusivity  $\mathcal{D}$ as
\begin{equation}
\begin{split}
\mathcal{D}   & =  \lim_{t\rightarrow\infty}\frac{(\xi \mu)^{2}\langle |\boldsymbol{F}^a|^{2} \rangle}{2}\left[\frac{-1}{\xi}\exp(-t/\xi)+\frac{1}{\xi}\right] \\
&= \frac{\xi \mu^{2}\langle |\boldsymbol{F}^a|^{2} \rangle }{2}.   \label{eq:diffusion0}
\end{split}
\end{equation}
Calculation of $\mathcal{D}$ using this formula requires the determination of the net active force autocorrelation time scale $\xi$ and variance $\langle |\boldsymbol{F}^a|^{2} \rangle$. Next, we explain how these two quantities can be estimated in some limits.


\subsection{Time-autocorrelation constant \label{sec:zetatheory}}

The autocorrelation time scale $\xi$ is a function of $Pe$, $\rho$ and $\mu$, and can be extracted numerically from the data of Fig.~\ref{force_autocorrelation_time_constant}b. Its dependence on P\'eclet number is illustrated in Fig.~\ref{force_autocorrelation_time_constant}c($i$). Consistent with the observations of panels a and b, we find that the correlation time increases with $Pe$. At low P\'eclet numbers, that increase is nearly linear; indeed, at low $Pe$, the typical duration of a collision scales with $Pe$ as we will explain in more detail below. At high P\'eclet numbers, the correlation time plateaus and asymptotes to a value slightly above 1; indeed, in that regime, most swimmers slide off the colloid before the end of their run, so that the typical duration of a collision is of order 1. The dependence of $\xi$ on density $\rho$ and mobility ratio $\mu$ is shown in Fig.~\ref{force_autocorrelation_time_constant}c($ii$): the correlation time is found to be roughly independent of $\mu$ and shows a slight decrease with $\rho$, albeit very weak.

A lower-bound estimate for the time constant $\xi$ can be derived analytically as the average time $\langle \tau_{int}\rangle$ that a swimmer interacts with the colloid over the course of one collision. This average time can be calculated exactly in the limit of low mobility ratios ($\mu\to 0$) based on the model of \citet{saintillan2023dispersion}. When a collision occurs with incidence angle $\alpha_0$, there are two possible outcomes: either the swimmer will finish its current run on the colloid surface, or it will escape the surface by sliding off tangentially. This outcome depends on the remaining time $\tau_r$ in the run after collision. In dimensionless variables, the time to escape tangentially, in the case of a fixed colloid ($\mu=0$), can be calculated as
\begin{equation}
    t_e(\alpha_0)=-\log \tan(\alpha_0/2).
\end{equation}
Escape will only occur if the remaining time $\tau_r$ exceeds the escape time $t_e$, which defines a critical incidence angle for escape, $\alpha_c(\tau)=2\tan^{-1}[\exp(-\tau_r)]$ (see Saintillan \cite{saintillan2023dispersion} for more details). Note that the collision time $\tau_r$ is a random variable over the interval $[0,\tau]$; for the purpose of estimating the longest possible interaction time during a run, we set $\tau_r=\tau$. The interaction time in that case is
\begin{equation}
  \tau_{int}(\alpha_0,\tau) =
    \begin{cases}
      \tau &\text{if $0 \leq \alpha_0 < \alpha_{c}(\tau)$},\\    
      t_{{e}} & \text{if $ \alpha_{c}(\tau) \leq \alpha_0 \leq \pi/2 $}.\\
    \end{cases}       
\end{equation}
In this expression, both $\alpha_0$ and $\tau$ are random variables. We first take an average over the incidence angle, which is uniformly distributed over $[0,\pi/2]$ \cite{saintillan2023dispersion}, yielding
\begin{equation}
   \langle \tau_{int} \rangle_{\alpha_0}(\tau) = \frac{2}{\pi}\tau\alpha_{c}(\tau)  -\frac{2}{\pi}\int_{\alpha_{c}(\tau)}^{\pi/2} \log \tan(\alpha_0/2)\, \mathrm{d} \alpha_0.
\end{equation}
Next, we integrate over exponentially distributed run times, providing the estimate for the time autocorrelation constant as:
\begin{equation}
    \xi = \langle \tau_{int} \rangle_{\alpha_0,\tau}=\int_{0}^{\infty} \langle \tau_{int} \rangle_{\alpha_0}(\tau)\frac{ \exp(-\tau/Pe)}{Pe}\, \mathrm{d} \tau. \label{eq:xiestimate}
\end{equation}
This estimate is plotted in Fig.~\ref{force_autocorrelation_time_constant}c($i$), where it is found to underpredict the correlation time at low $Pe$, but captures the correct asymptote at high $Pe$.\ The departure at low $Pe$ can be rationalized as follows. When the P\'eclet number is small, most runs involving a collision end on the surface of the colloid; half of the subsequent runs then start with a collision, leading to a longer interaction time than estimated here. On the other hand, at high $Pe$, nearly all collisions result in an escape, which explains why the estimate of eqn~(\ref{eq:xiestimate}) matches the data well in that limit.



\subsection{Fluctuating active force variance}

We now seek to estimate the force variance  $\langle |\boldsymbol{F}^a|^{2} \rangle$, where the net active force was defined in eqn~(\ref{eq:activeforce}). Here again, we focus on the limit of low-mobility ratio, $\mu\to 0$. In that limit, the contact force magnitudes in eqn~(\ref{eq:Fsystem}) become decoupled, and are simply given by
\begin{equation}
    F_i^c(t) = \boldsymbol{p}_i(t)\cdot \boldsymbol{q}_i(t)=-\cos \alpha_i(t).
\end{equation}
The net instantaneous active force on the colloid resulting from swimmer collisions is therefore 
\begin{equation}
 \boldsymbol{F}^a(t) = -\sum_{i=1}^{c(t)}  \cos \alpha_{i}(t)  \bigg(\begin{matrix}
\cos \beta_{i}(t)\\
\sin \beta_{i}(t)
\end{matrix}\bigg),
\end{equation}
and its variance can be obtained as
\begin{equation}
\langle |\boldsymbol{F}^{a}|^{2} \rangle
=  \bigg \langle \sum_{i=1}^{c(t)} \sum_{j=1}^{c(t)} \Theta (\alpha_{i},\alpha_{j}) \cdot \Xi (\beta_{i},\beta_{j})
\bigg \rangle,    
\end{equation}
where $\Theta (\alpha_{i},\alpha_{j}) = \cos \alpha_{i}(t) \cdot \cos \alpha_{j}(t)$ and $\Xi(\beta_{i},\beta_{j}) = \cos \beta_{i}(t) \cdot\cos \beta_{j}(t)  + \sin \beta_{i}(t)\cdot\sin \beta_{j}(t)$. The ensemble average is over all possible values of $c(t)$, $\alpha_i$ and $\beta_i$. The angular positions $\beta_i$ of the swimmers on the colloid surface are uniformly distributed between $0$ and $2\pi$, and are uncorrelated. Averaging over $\beta_i$ and $\beta_j$ therefore gives
\begin{equation}
    \int_0^{2\pi}\int_0^{2\pi} \Xi(\beta_{i},\beta_{j}) p(\beta_i)p(\beta_j)\,\mathrm{d}\beta_i\, \mathrm{d}\beta_j=\delta_{ij},
\end{equation}
so that the variance becomes
\begin{equation}
    \langle |\boldsymbol{F}^{a}|^{2} \rangle=\bigg \langle \sum_{i=1}^{c(t)} \sum_{j=1}^{c(t)} \Theta (\alpha_{i},\alpha_{j})\delta_{ij} \bigg \rangle = \bigg \langle \sum_{i=1}^{c(t)} \cos^2\alpha_i(t) \bigg \rangle.     
\end{equation}
In the low-mobility-ratio limit, the dynamics of the swimmers on the surface are uncorrelated, therefore the variance further simplifies to
\begin{equation}
\langle |\boldsymbol{F}^{a}|^{2} \rangle=   \langle c \rangle \langle \cos ^2 \alpha\rangle ,  \label{eq:varforce}
\end{equation}
where $\langle c\rangle$ is the average number of swimmers interacting with the colloid at any given time, and $\langle \cos^2\alpha\rangle$ is the average squared cosine of the contact angle. Next, we turn to the determination of these two quantities. 

\subsubsection{Surface swimmer density \label{sec:surface density}}

The mean number of swimmers on the surface of the colloid can also be estimated based on the prior work of \citet{saintillan2023dispersion} for a fixed colloid. As explained in that model, swimmer-colloid collisions are of two types: collisions of type A involve a swimmer starting its run in the bulk and encountering the colloid, while collisions of type B involve a swimmer starting its run on the colloid surface after a tumble leading to a new orientation pointing towards the colloid. For a collision of type B to occur, the swimmer should had ended its previous run on the colloid surface. As shown by \citet{saintillan2023dispersion}, the probabilities for any swimmer to undergo a collision of type A or B during any given run can be calculated as
\begin{equation}
P^{c}_{\!\mathrm{A}} = 1 - \exp\bigg(-\frac{2Pe}{\pi}\phi\bigg), \quad 
P^{c}_{\!\mathrm{B}} =\bigg(\frac{1- P^{esc}_{\!\mathrm{A}}}{1 + P^{esc}_{\!\mathrm{B}}}\bigg)P^{c}_{\!\mathrm{A}},  \label{eq:collisionprobs}
\end{equation}
where $\phi = \pi/ L^2$ is the area fraction of the colloid in the square domain of size $L$, and $P^{esc}_{\!\mathrm{A}}$ and $P^{esc}_{\!\mathrm{B}}$ are the probabilities for a collision of type A or B to lead to an escape from the pillar surface. These escape probabilities can be obtained by analysis of collision dynamics and were found to be 
\begin{equation}
    P^{esc}_{\!\mathrm{A}} = 1+\frac{1}{Pe}\left(\bar{\alpha}-\frac{\pi}{2} \right), \quad P^{esc}_{\!\mathrm{B}}=1-\frac{2}{\pi}\bar{\alpha}, \label{eq:escapeprobs}
\end{equation}
with $\bar{\alpha}=2\tan^{-1}[\exp(-Pe)]$. Note that eqn~(\ref{eq:collisionprobs}) and (\ref{eq:escapeprobs}) were technically derived for constant run time \cite{saintillan2023dispersion}, here taken equal to $Pe$; we will show below that they yield a quantitative estimate for $\langle c\rangle$ despite this approximation. 

\begin{figure}[t]
\centering
\includegraphics[width=0.40\textwidth]{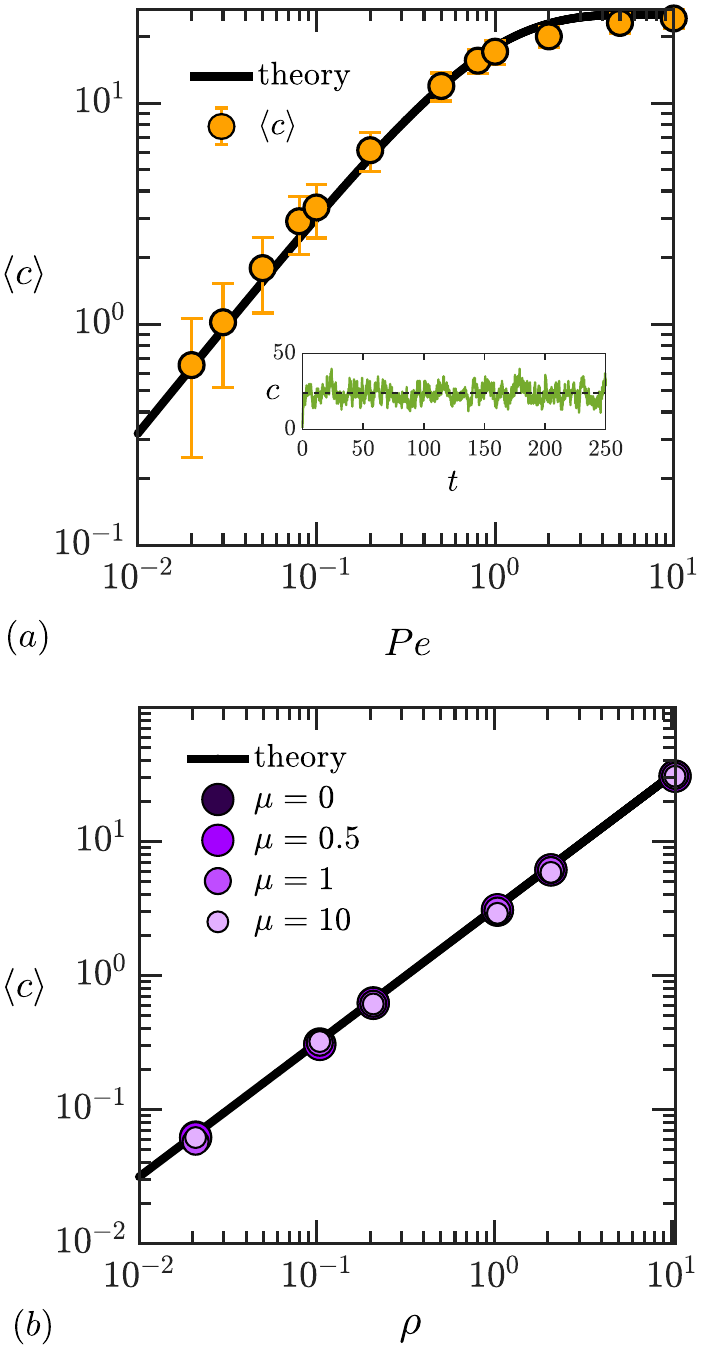}\vspace{-0.2cm}
\caption{Dependence of the mean number $\langle c \rangle$ of swimmers in contact with the colloid with (a) P\'eclet number $Pe$, for $\rho=8.04$ and $\mu=0.5$; and (b) density of the swimmers $\rho$ for $Pe=8.0$ and different mobility ratios. In (a), the error bars show half a standard deviation around the mean, and the inset illustrates temporal variations of $c(t)$ at $Pe=10$. In both panels, the solid black curves show the theoretical prediction of eqn~(\ref{eq:swimmers_in_contact}) .}\vspace{-0.0cm}
\label{fig_mean_number_swimmers}
\end{figure}

We can now estimate the probability $P^s$ that a swimmer will end its current run on the surface of the colloid. For this to happen, the swimmer must undergo a collision of either type A or B and finish its run on the surface. Therefore,
\begin{align}
\begin{split}
    P^s&=P^{c}_{\!\mathrm{A}}(1-P^{esc}_{\!\mathrm{A}})+P^{c}_{\!\mathrm{B}}(1-P^{esc}_{\!\mathrm{B}}) \\
    &=2 \bigg(\frac{1- P^{esc}_{\!\mathrm{A}}}{1 + P^{esc}_{\!\mathrm{B}}}\bigg)\left[1 - \exp\bigg(-\frac{2Pe}{L^2}\bigg)\right].   \label{eq:Ps}
\end{split}
\end{align}
The mean number of swimmers in contact with the colloid can then be estimated as $\langle c \rangle = \nu P^s$, where $\nu$ is the total number of swimmers in the domain. In the limit of $L\gg 1$ (large domain size), we can expand the exponential in a Taylor series, yielding the simple form
\begin{equation}
    \langle c \rangle \approx \frac{\pi^2-4\pi\tan^{-1}[\exp(-Pe)]}{\pi-2\tan^{-1}[\exp(-Pe)]}\rho, \label{eq:swimmers_in_contact}
\end{equation}
where $\rho$ is the mean swimmer number density introduced in eqn~(\ref{eq:rho}). The linear dependence on $\rho$ is unsurprising, since swimmers are non-interacting in our model. At low P\'eclet number, the number of swimmers on the surface grows linearly with $Pe$ as $\langle c \rangle \approx 4 Pe \rho$: indeed, in that regime, most swimmers incurring a collision will finish their run on the colloid surface and therefore spend a time on the surface that scales linearly with $Pe$. On the other hand, at high P\'eclet number, $\langle c \rangle$ becomes independent of $Pe$ and asymptotes to $\pi \rho$: in this regime, most swimmers colliding with the colloid will escape the colloid surface before the end of their run, and therefore spend a time on the colloid surface that is independent of $Pe$. 


We test the prediction of eqn~(\ref{eq:swimmers_in_contact}) against numerical data in Fig.~\ref{fig_mean_number_swimmers}. The dependence on $Pe$ is shown in panel a for a density of $\rho=8.04$ and a mobility ratio of $\mu=0.5$, and excellent agreement is found between the theory and the numerical data, despite the approximations made (constant run length, fixed colloid). In particular, the low- and high-P\'eclet-number asymptotes match the data perfectly; the largest deviations are observed for $Pe=O(1)$, where the agreement remains very good. The inset displays the temporal variation of the number of swimmers in contact $c(t)$ with the colloid for $Pe=10$ and highlights significant variations around the mean. The variance of $c(t)$ is in fact found to be equal to the mean, $\mathrm{Var}(c) = \langle c \rangle$, consistent with Poisson statistics.  Panel b shows the dependence of $\langle c \rangle$ with $\rho$ for $Pe=8.0$ and for different mobility ratios. Here again, excellent match is obtained between theory and data, independent of the value of $\mu$. 

\begin{figure}[t]
\includegraphics[width=0.52\textwidth]{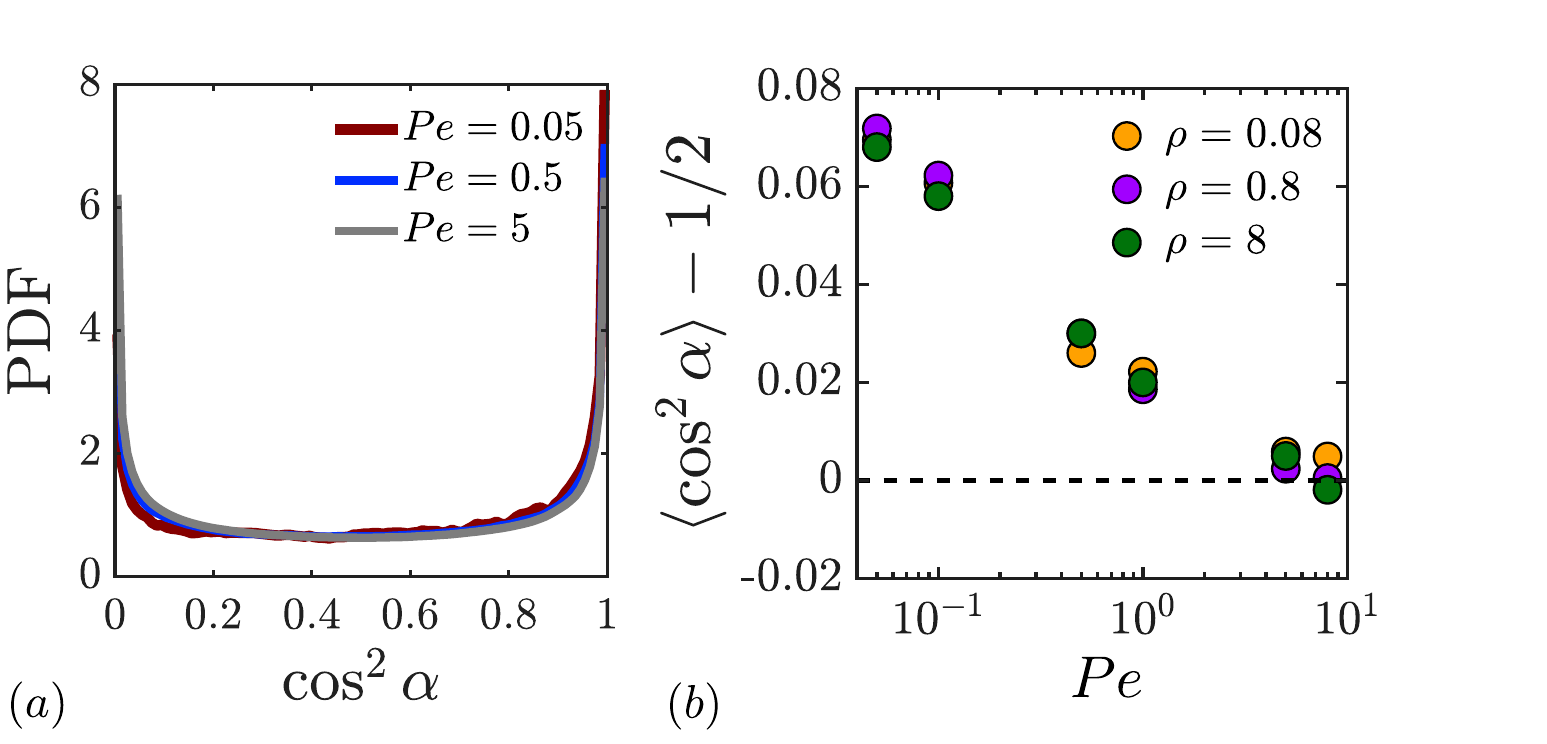}
\caption{(a) Probability density functions of the squared cosine of the contact angle $\alpha$ for different P\'eclet numbers $Pe$,  obtained from numerical simulations with $\rho= 8.04$ and $\mu= 0.01$. (b) Deviation of $\langle \cos^{2}\alpha \rangle$ about 1/2 as a function of P\'eclet number for different swimmer densities and $\mu=0.01$.}
\label{histogram_incidence_angles}
\end{figure}

\subsubsection{Contact angle distribution \label{sec:contactangle}}

To complete the calculation of the force variance, we need an estimate for the average squared cosine of the contact angle $\alpha$. The probability density function for $\alpha$ is not easily obtained theoretically: indeed, it depends on the distribution of the incidence angle $\alpha_0$ at the start of a collision, as well as on the dynamics of the swimmer on the colloid surface during contact and on any possible tumbles that may occur during the collision. In particular, we expect a dependence on $Pe$. To estimate $\langle \cos^2\alpha\rangle $, we instead turn to numerical data, and plot in Fig.~\ref{histogram_incidence_angles}a the probability density function of  $ \cos^{2}\alpha$ for different P\'eclet numbers. The probability density functions peak near $\cos^2\alpha=0$ and $1$, and are nearly symmetric about $1/2$, with a slight bias towards $1$ that is very weak and most visible at low values of $Pe$. As a result, we expect $\langle \cos^2\alpha\rangle \approx 1/2$ over a wide range of $Pe$. Departures from this value are quantified more precisely in Fig.~\ref{histogram_incidence_angles}b, where they are most significant at low $Pe$ but never exceed 10\% over the whole range of P\'eclet numbers considered in this study. 

\begin{figure}[t]
\centering
\includegraphics[width=0.42\textwidth]{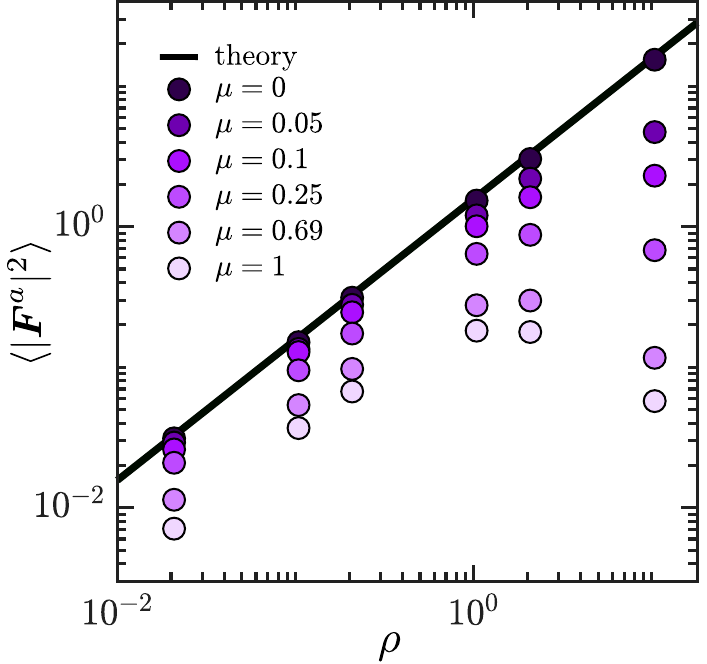}
\caption{Variance $\langle |\boldsymbol{F}^a|^{2}\rangle$ of the net active force on the passive colloid as a function of swimmer number density $\rho$ for different mobility ratios $\mu$ for $Pe=8$. The solid black line shows the analytical expression of eqn~(\ref{eq:varforce}).}
\label{force_variance}
\end{figure}

\subsubsection{Force variance}

\noindent Figure \ref{force_variance} shows the active force variance $\langle |\boldsymbol{F}^{a}|^{2} \rangle$ as a function of swimmer number density $\rho$, for a P\'eclet number of $Pe=8$. The plot compares numerical data for different values of the mobility ratio with the derived theoretical estimate of $\langle c \rangle \langle \cos^{2} \alpha \rangle$, which is technically valid in the limit of $\mu\to 0$. At zero mobility ratio, there is a quantitative agreement between data and theory. As the mobility ratio increases, the force variance is found to decrease; the dependence on $\rho$ remains linear at low densities, but significant departures are observed for moderate values of $\mu$ at high densities. Since our estimates for both $\langle c\rangle$ and $\langle \cos^2\alpha\rangle$ are valid independent of $\rho$ and $\mu$ (see Sec.~\ref{sec:surface density} and \ref{sec:contactangle}), this suggests that it is the derivation of eqn~(\ref{eq:varforce}) itself that breaks down. Indeed, when deriving eqn~(\ref{eq:varforce}), we assumed $\mu=0$, which entirely decouples the dynamics of the swimmers on the colloid surface as can be seen from eqn~(\ref{eq:Fsystem}). At moderate mobility ratios, correlations between swimmers will affect the contact forces, and these correlations should become all the more significant at high densities when many swimmers interact with the colloid at any given time. In particular, the decrease in $\langle |\boldsymbol{F}^{a}|^{2} \rangle$ seen in Fig.~\ref{force_variance} suggests that correlations may result in a partial cancelation of forces, which is not captured by our simple theoretical model.


\subsection{Effective dispersion}

Combining the results of eqn~(\ref{eq:diffusion0}), (\ref{eq:varforce}) and (\ref{eq:swimmers_in_contact}), we obtain the following theoretical prediction for the long-time  diffusivity:
\begin{equation}
\mathcal{D} \approx \bigg(\frac{\pi^{2} -4\pi \tan^{-1}[\exp(-Pe)]}{\pi -2 \tan^{-1}[\exp(-Pe)]}\bigg)\frac{\xi(Pe) \mu^{2}\rho}{4} . \label{eq:diffusivity_final}
\end{equation}
The correlation time $\xi(Pe)$ can either be obtained numerically from simulation data as in Fig.~\ref{force_autocorrelation_time_constant}, or estimated theoretically using eqn~(\ref{eq:xiestimate}), with the caveat that this estimate is quantitative only at high P\'eclet number. In that limit ($Pe\to \infty$), eqn~(\ref{eq:diffusivity_final}) can be expanded to obtain the asymptotic result,
\begin{equation}
    \mathcal{D}\approx 0.91\mu^2\rho,
\end{equation}
or, in dimensional variables,
\begin{equation}
    \mathcal{D}\approx 0.91\mu^2\rho u A^3. \label{eq:highPe}
\end{equation}

\begin{figure}[t]
\centering
\includegraphics[width=0.4\textwidth]{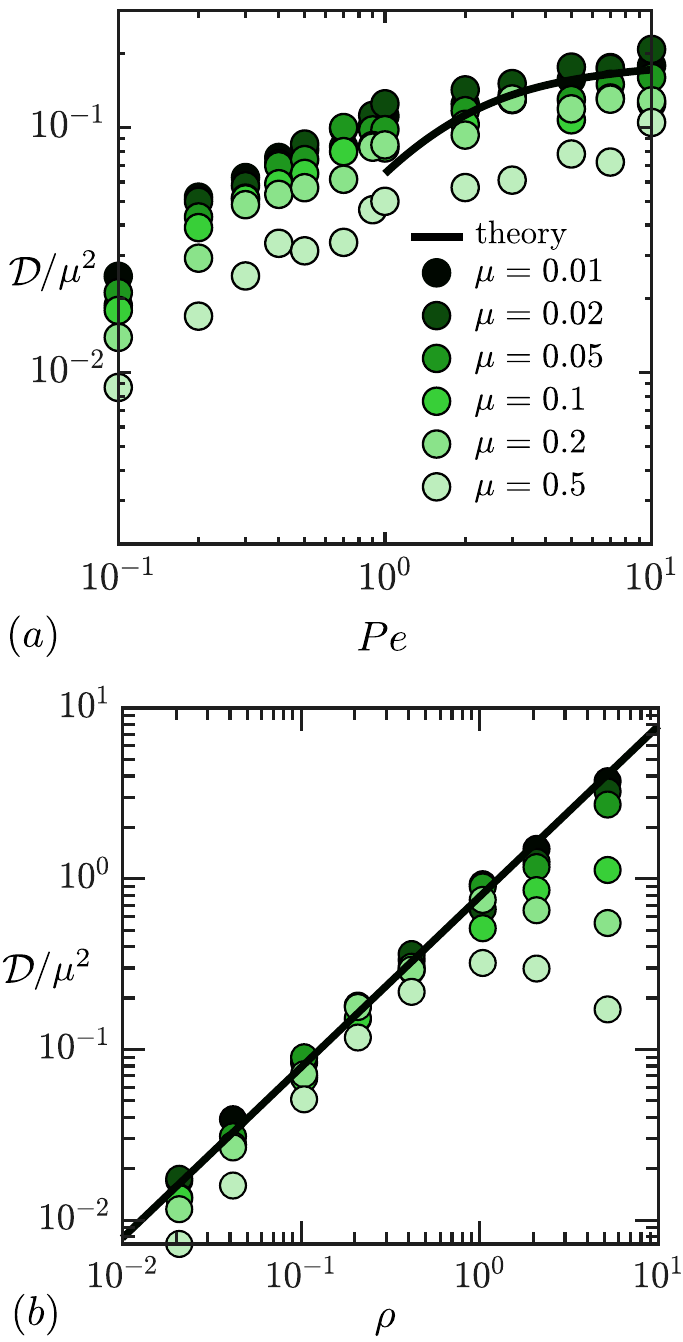}
\caption{Variation of the diffusivity $\mathcal{D}$, scaled by the square of the mobility ratio $\mu^{2}$, with: (a) P\'eclet number $Pe$, for a fixed swimmer density of $\rho = 0.2$; and (b) swimmer density $\rho$ for $Pe=8$. Symbols show results from stochastic simulations for different values of $\mu$, while the solid black curves show the semi-analytical model of eqn~(\ref{eq:diffusivity_final}).}
\label{Fig_diffusivity_variations_2}
\end{figure}

We compare the prediction of eqn~(\ref{eq:diffusivity_final}) to simulation data in Fig.~\ref{Fig_diffusivity_variations_2}, showing the variation of the effective diffusivity $\mathcal{D}$, scaled by $\mu^2$, with P\'eclet number and swimmer density. The theoretical prediction in Fig.~\ref{Fig_diffusivity_variations_2} uses the analytical estimate of eqn~(\ref{eq:xiestimate}) for $\xi(Pe)$. In both plots, rescaling $\mathcal{D}$ by $\mu^2$ collapses the data at low mobility ratios in agreement with the model, with some departures observed for $\mu\gtrsim 0.25$, especially at large densities, consistent with the observations of Figs.~\ref{Fig_diffusivity_variations}c and \ref{force_variance}.
Figure \ref{Fig_diffusivity_variations_2}a shows the variation of $\mathcal{D}/\mu^2$ with P\'eclet number. The theoretical model captures the data very well in the limit of $Pe \geq 1$ and $\mu \longrightarrow 0$. At lower P\'eclet numbers, the theory departs from the simulation results as expected, and could be improved by deriving more accurate estimates for $\xi(Pe)$ and $\langle \cos^2\alpha\rangle$. 
The dependence on swimmer density $\rho$ is investigated in Fig. \ref{Fig_diffusivity_variations_2}b, where the linear prediction of eqn~(\ref{Fig_diffusivity_variations_2}) captures the data quantitatively at low mobility ratios and low to moderate densities. As expected based on the model assumptions, departures are observed at high $\mu$ and $\rho$ and can be traced back to departures in the force variance in Fig.~\ref{force_variance}.

In summary, our theory agrees very well with particle simulations in the regimes of low swimmer density, low mobility ratio, and high P\'eclet numbers, and provides a simple analytical framework for the description of long-time colloidal dispersion in baths of run-and-tumble microswimmers. It should be noted that the parameter regime in which our model works best is experimentally relevant. Using an experimental estimate of  $\zeta \sim 10^{-8} - 10^{-7}\,$Ns/m for the drag on a bacterium \cite{chattopadhyay2006swimming,lagarde2020colloidal}, along with Stokes drag $Z = 1/M = 6\pi\eta A$ for a colloid in water with radius $A=0.5-10\,\mu$m \cite{mino2011enhanced,angelani2011effective,lagarde2020colloidal,grober2023unconventional},  we estimate the mobility ratio to be in the range of $\mu \in [0.1,10]$ in typical experiments. Yet lower mobility ratios, reaching down to $0.01$, occur in systems involving colloidal clusters \cite{grober2023unconventional}.
Based on experimental estimates of typical bacterial swim speeds ($u=5-25\,\mu$m/sec) \cite{chattopadhyay2006swimming,mino2011enhanced,lagarde2020colloidal,grober2023unconventional} and mean run time ($\bar{\tau}\sim 1$ sec) \cite{grober2023unconventional}, as well as colloid sizes, we estimate the P\'eclet number to vary in the range of  $Pe\in[1, 50]$. Finally, the typical 2D density of swimmer suspensions, based on recent experiments near flat substrates  \cite{lagarde2020colloidal,grober2023unconventional}, is in the range of $\rho \in [0.01, 1]$.



We also note that our model, and in particular the predicted linear dependence on $\rho$ and $u$ at high P\'eclet numbers, is consistent with various past experiments and models. In early experiments using \textit{C. reinhardtii}, \citet{leptos2009dynamics} experimentally measured the diffusivity of suspended tracers and observed a clear linear dependence with number density. 
Subsequent experiments performed by \citet{mino2011enhanced} on the diffusion of passive particles near a solid substrate in \textit{E. coli} suspensions found a linear dependence on the active flux $J=\rho_v u$, where $\rho_v$ is the volume density of bacteria. This result again agrees with eqn~(\ref{eq:highPe}), although they found only a weak dependence on the size of the colloid, which they attributed to a hydrodynamic mechanism for the diffusivity enhancement. A similar result was obtained \citet{jepson2013enhanced} for the diffusion of non-motile bacteria in a motile bath. Hydrodynamic theories such as those of \citet{lin2011stirring} or \citet{kasyap2014hydrodynamic}, which explain the diffusivity enhancement based on dipolar flow disturbances induced by microswimmers, indeed predict a linear dependence on the active flux $J$, but with a prefactor that is independent of $\mu$ and depends on the fourth power of microswimmer size rather than colloid size. \citet{lagarde2020colloidal} performed both experiments and numerical simulations based on steric interactions; they found a linear dependence of $\mathcal{D}$ on $u$, but did not characterize the dependence on swimmer density and mobility ratio. More recently, \citet{burkholder2017tracer} derived a model for the diffusivity of a passive colloid resulting from steric interactions with a bath of active Brownian particles, and predicted a quadratic dependence on $\mu$ as well as a linear dependence on $\rho u$ in the limit of persistent swimmers. They also predicted a quadratic dependence on $u$ at low P\'eclet number, which is also consistent with our data in Fig.~\ref{Fig_diffusivity_variations}b. To the best of our knowledge, our model is the first to specifically address the case of athermal run-and-tumble swimmers, and to provide a semi-analytical prediction for the dependence of $\mathcal{D}$ on all system parameters for that particular system.

\section{Conclusions}\label{conclusions}

Using stochastic simulations and an asymptotic theory, we have analyzed the dispersion of a passive colloid immersed in an unbounded suspension of non-interacting active run-and-tumble particles in two dimensions. We found a linear increase of the colloid diffusivity with swimmer density at low to moderate densities, consistent with past experiments and models \citep{leptos2009dynamics,mino2011enhanced, jepson2013enhanced,morozov2014enhanced, burkholder2017tracer, grober2023unconventional}, followed by a peak and slight decrease at very high densities, in a regime that is not experimentally relevant. The variation of $\mathcal{D}$ with the persistence of the swimmer trajectories, as captured by the P\'eclet number, shows a linear increase with $Pe$ that slows down and plateaus for $Pe\gtrsim 1$; in dimensional variables, that corresponds to a diffusivity that scales quadratically then linearly with the microswimmer speed $u$, once again in agreement with past models \cite{burkholder2017tracer}. The dependence of $\mathcal{D}$ on the mobility ratio of the colloid to the swimmer was found to be quadratic at low mobility ratios but to saturate for $\mu\gtrsim 1$ (small colloids). 

Using a theoretical model based on the Green-Kubo formula, we related the long-time colloid diffusivity to the variance of the net fluctuating force on the colloid resulting from collisions with the swimmers. A semi-analytical expression for the force variance was obtained under the assumptions of low mobility ratio ($\mu\lesssim 0.25$), low swimmer density ($\rho\lesssim 1$), and high P\'eclet number ($Pe\gtrsim 1$), a regime relevant to many experimental systems. In that regime, the dependence of $\mathcal{D}$ on system parameters is simple and given by eqn~(\ref{eq:highPe}), and quantitative agreement is found between theory and stochastic simulations.


The model introduced herein remains minimal and neglects 
many complexities inherent to biological and synthetic active systems \citep{volpe2011microswimmers, kantsler2013ciliary, molaei2014failed}. For instance, several experiments demonstrate that biochemical signaling 
\citep{korobkova2004molecular,korobkova2006hidden,wang2017non} and the large variability of run times of the \textit{E. Coli} cells \cite{figueroa20203d} can lead to a non-Poissonian distribution for the run-and-tumble statistics. Such distributions could be added to our current model with relative ease. Many experiments involve a monolayer of colloids sedimented on a flat substrate in a bacterial suspension. In these systems, hydrodynamic interactions with the wall can have various effects including enhanced drag, bacterial accumulation at the wall \citep{berke2008hydrodynamic,ezhilan2015transport}, swimming in circles \citep{drescher2011fluid, lauga2006swimming, frymier1995three}. In some cases, bacteria navigate either above or below the colloid \cite{lagarde2020colloidal}, an effect that could be captured by extending the model to three dimensions. Such extension to 3D was recently discussed by \citet{saintillan2023dispersion}, and may allow further studies of colloid sedimentation in active suspensions \cite{singh2021bacterial,maldonado2023sedimentation}, or of melting of colloidal crystals by active baths \cite{massana2024multiple}. Other effects that are of interest are hydrodynamic interactions between the swimmers and colloid \cite{lin2011stirring,kasyap2014hydrodynamic}, which can significantly alter scattering dynamics \cite{spagnolie2015geometric}, as well as the effect of swimmer \citep{sipos2015hydrodynamic, wysocki2015giant} and colloid \citep{di2010bacterial, sokolov2010swimming, peng2016diffusion, baek2018generic} shape. Some of these effects are straightforward extensions of the present model and may be addressed in future work.

\section*{Acknowledgements}
The authors gratefully acknowledge discussions with J\'er\'emie Palacci and Daniel Grober, as well as
funding from National Science Foundation Grant No.\ CBET-1934199 (D.S.). 


\bibliography{main}

\begin{thebibliography}{76}%
\makeatletter
\providecommand \@ifxundefined [1]{%
 \@ifx{#1\undefined}
}%
\providecommand \@ifnum [1]{%
 \ifnum #1\expandafter \@firstoftwo
 \else \expandafter \@secondoftwo
 \fi
}%
\providecommand \@ifx [1]{%
 \ifx #1\expandafter \@firstoftwo
 \else \expandafter \@secondoftwo
 \fi
}%
\providecommand \natexlab [1]{#1}%
\providecommand \enquote  [1]{``#1''}%
\providecommand \bibnamefont  [1]{#1}%
\providecommand \bibfnamefont [1]{#1}%
\providecommand \citenamefont [1]{#1}%
\providecommand \href@noop [0]{\@secondoftwo}%
\providecommand \href [0]{\begingroup \@sanitize@url \@href}%
\providecommand \@href[1]{\@@startlink{#1}\@@href}%
\providecommand \@@href[1]{\endgroup#1\@@endlink}%
\providecommand \@sanitize@url [0]{\catcode `\\12\catcode `\$12\catcode
  `\&12\catcode `\#12\catcode `\^12\catcode `\_12\catcode `\%12\relax}%
\providecommand \@@startlink[1]{}%
\providecommand \@@endlink[0]{}%
\providecommand \url  [0]{\begingroup\@sanitize@url \@url }%
\providecommand \@url [1]{\endgroup\@href {#1}{\urlprefix }}%
\providecommand \urlprefix  [0]{URL }%
\providecommand \Eprint [0]{\href }%
\providecommand \doibase [0]{http://dx.doi.org/}%
\providecommand \selectlanguage [0]{\@gobble}%
\providecommand \bibinfo  [0]{\@secondoftwo}%
\providecommand \bibfield  [0]{\@secondoftwo}%
\providecommand \translation [1]{[#1]}%
\providecommand \BibitemOpen [0]{}%
\providecommand \bibitemStop [0]{}%
\providecommand \bibitemNoStop [0]{.\EOS\space}%
\providecommand \EOS [0]{\spacefactor3000\relax}%
\providecommand \BibitemShut  [1]{\csname bibitem#1\endcsname}%
\let\auto@bib@innerbib\@empty
\bibitem [{\citenamefont {Wu}\ and\ \citenamefont
  {Libchaber}(2000)}]{wu2000particle}%
  \BibitemOpen
  \bibfield  {author} {\bibinfo {author} {\bibfnamefont {Xiao-Lun}\
  \bibnamefont {Wu}}\ and\ \bibinfo {author} {\bibfnamefont {Albert}\
  \bibnamefont {Libchaber}},\ }\bibfield  {title} {\enquote {\bibinfo {title}
  {Particle diffusion in a quasi-two-dimensional bacterial bath},}\ }\href@noop
  {} {\bibfield  {journal} {\bibinfo  {journal} {Phys. Rev. Lett.}\ }\textbf
  {\bibinfo {volume} {84}},\ \bibinfo {pages} {3017} (\bibinfo {year}
  {2000})}\BibitemShut {NoStop}%
\bibitem [{\citenamefont {Hernandez-Ortiz}\ \emph {et~al.}(2005)\citenamefont
  {Hernandez-Ortiz}, \citenamefont {Stoltz},\ and\ \citenamefont
  {Graham}}]{hernandez2005transport}%
  \BibitemOpen
  \bibfield  {author} {\bibinfo {author} {\bibfnamefont {Juan~P}\ \bibnamefont
  {Hernandez-Ortiz}}, \bibinfo {author} {\bibfnamefont {Christopher~G}\
  \bibnamefont {Stoltz}}, \ and\ \bibinfo {author} {\bibfnamefont {Michael~D}\
  \bibnamefont {Graham}},\ }\bibfield  {title} {\enquote {\bibinfo {title}
  {Transport and collective dynamics in suspensions of confined swimming
  particles},}\ }\href@noop {} {\bibfield  {journal} {\bibinfo  {journal}
  {Phys. Rev. Lett.}\ }\textbf {\bibinfo {volume} {95}},\ \bibinfo {pages}
  {204501} (\bibinfo {year} {2005})}\BibitemShut {NoStop}%
\bibitem [{\citenamefont {Mi{\~n}o}\ \emph {et~al.}(2011)\citenamefont
  {Mi{\~n}o}, \citenamefont {Mallouk}, \citenamefont {Darnige}, \citenamefont
  {Hoyos}, \citenamefont {Dauchet}, \citenamefont {Dunstan}, \citenamefont
  {Soto}, \citenamefont {Wang}, \citenamefont {Rousselet},\ and\ \citenamefont
  {Clement}}]{mino2011enhanced}%
  \BibitemOpen
  \bibfield  {author} {\bibinfo {author} {\bibfnamefont {Gast{\'o}n}\
  \bibnamefont {Mi{\~n}o}}, \bibinfo {author} {\bibfnamefont {Thomas~E}\
  \bibnamefont {Mallouk}}, \bibinfo {author} {\bibfnamefont {Thierry}\
  \bibnamefont {Darnige}}, \bibinfo {author} {\bibfnamefont {Mauricio}\
  \bibnamefont {Hoyos}}, \bibinfo {author} {\bibfnamefont {Jeremi}\
  \bibnamefont {Dauchet}}, \bibinfo {author} {\bibfnamefont {Jocelyn}\
  \bibnamefont {Dunstan}}, \bibinfo {author} {\bibfnamefont {Rodrigo}\
  \bibnamefont {Soto}}, \bibinfo {author} {\bibfnamefont {Yang}\ \bibnamefont
  {Wang}}, \bibinfo {author} {\bibfnamefont {Annie}\ \bibnamefont {Rousselet}},
  \ and\ \bibinfo {author} {\bibfnamefont {Eric}\ \bibnamefont {Clement}},\
  }\bibfield  {title} {\enquote {\bibinfo {title} {Enhanced diffusion due to
  active swimmers at a solid surface},}\ }\href@noop {} {\bibfield  {journal}
  {\bibinfo  {journal} {Phys. Rev. Lett.}\ }\textbf {\bibinfo {volume} {106}},\
  \bibinfo {pages} {048102} (\bibinfo {year} {2011})}\BibitemShut {NoStop}%
\bibitem [{\citenamefont {Grober}\ \emph {et~al.}(2023)\citenamefont {Grober},
  \citenamefont {Palaia}, \citenamefont {U{\c{c}}ar}, \citenamefont {Hannezo},
  \citenamefont {{\v{S}}ari{\'c}},\ and\ \citenamefont
  {Palacci}}]{grober2023unconventional}%
  \BibitemOpen
  \bibfield  {author} {\bibinfo {author} {\bibfnamefont {Daniel}\ \bibnamefont
  {Grober}}, \bibinfo {author} {\bibfnamefont {Ivan}\ \bibnamefont {Palaia}},
  \bibinfo {author} {\bibfnamefont {Mehmet~Can}\ \bibnamefont {U{\c{c}}ar}},
  \bibinfo {author} {\bibfnamefont {Edouard}\ \bibnamefont {Hannezo}}, \bibinfo
  {author} {\bibfnamefont {Andela}\ \bibnamefont {{\v{S}}ari{\'c}}}, \ and\
  \bibinfo {author} {\bibfnamefont {J{\'e}r{\'e}mie}\ \bibnamefont {Palacci}},\
  }\bibfield  {title} {\enquote {\bibinfo {title} {Unconventional colloidal
  aggregation in chiral bacterial baths},}\ }\href@noop {} {\bibfield
  {journal} {\bibinfo  {journal} {Nat. Phys.}\ }\textbf {\bibinfo {volume}
  {19}},\ \bibinfo {pages} {1680--1688} (\bibinfo {year} {2023})}\BibitemShut
  {NoStop}%
\bibitem [{\citenamefont {Angelani}\ \emph {et~al.}(2009)\citenamefont
  {Angelani}, \citenamefont {Di~Leonardo},\ and\ \citenamefont
  {Ruocco}}]{angelani2009self}%
  \BibitemOpen
  \bibfield  {author} {\bibinfo {author} {\bibfnamefont {Luca}\ \bibnamefont
  {Angelani}}, \bibinfo {author} {\bibfnamefont {Roberto}\ \bibnamefont
  {Di~Leonardo}}, \ and\ \bibinfo {author} {\bibfnamefont {Giancarlo}\
  \bibnamefont {Ruocco}},\ }\bibfield  {title} {\enquote {\bibinfo {title}
  {Self-starting micromotors in a bacterial bath},}\ }\href@noop {} {\bibfield
  {journal} {\bibinfo  {journal} {Phys. Rev. Lett.}\ }\textbf {\bibinfo
  {volume} {102}},\ \bibinfo {pages} {048104} (\bibinfo {year}
  {2009})}\BibitemShut {NoStop}%
\bibitem [{\citenamefont {Di~Leonardo}\ \emph {et~al.}(2010)\citenamefont
  {Di~Leonardo}, \citenamefont {Angelani}, \citenamefont {Dell’Arciprete},
  \citenamefont {Ruocco}, \citenamefont {Iebba}, \citenamefont {Schippa},
  \citenamefont {Conte}, \citenamefont {Mecarini}, \citenamefont {De~Angelis},\
  and\ \citenamefont {Di~Fabrizio}}]{di2010bacterial}%
  \BibitemOpen
  \bibfield  {author} {\bibinfo {author} {\bibfnamefont {Roberto}\ \bibnamefont
  {Di~Leonardo}}, \bibinfo {author} {\bibfnamefont {Luca}\ \bibnamefont
  {Angelani}}, \bibinfo {author} {\bibfnamefont {Dario}\ \bibnamefont
  {Dell’Arciprete}}, \bibinfo {author} {\bibfnamefont {Giancarlo}\
  \bibnamefont {Ruocco}}, \bibinfo {author} {\bibfnamefont {Valerio}\
  \bibnamefont {Iebba}}, \bibinfo {author} {\bibfnamefont {Serena}\
  \bibnamefont {Schippa}}, \bibinfo {author} {\bibfnamefont {Maria~Pia}\
  \bibnamefont {Conte}}, \bibinfo {author} {\bibfnamefont {Francesco}\
  \bibnamefont {Mecarini}}, \bibinfo {author} {\bibfnamefont {Francesco}\
  \bibnamefont {De~Angelis}}, \ and\ \bibinfo {author} {\bibfnamefont {Enzo}\
  \bibnamefont {Di~Fabrizio}},\ }\bibfield  {title} {\enquote {\bibinfo {title}
  {Bacterial ratchet motors},}\ }\href@noop {} {\bibfield  {journal} {\bibinfo
  {journal} {PNAS}\ }\textbf {\bibinfo {volume} {107}},\ \bibinfo {pages}
  {9541--9545} (\bibinfo {year} {2010})}\BibitemShut {NoStop}%
\bibitem [{\citenamefont {Sokolov}\ \emph {et~al.}(2010)\citenamefont
  {Sokolov}, \citenamefont {Apodaca}, \citenamefont {Grzybowski},\ and\
  \citenamefont {Aranson}}]{sokolov2010swimming}%
  \BibitemOpen
  \bibfield  {author} {\bibinfo {author} {\bibfnamefont {Andrey}\ \bibnamefont
  {Sokolov}}, \bibinfo {author} {\bibfnamefont {Mario~M}\ \bibnamefont
  {Apodaca}}, \bibinfo {author} {\bibfnamefont {Bartosz~A}\ \bibnamefont
  {Grzybowski}}, \ and\ \bibinfo {author} {\bibfnamefont {Igor~S}\ \bibnamefont
  {Aranson}},\ }\bibfield  {title} {\enquote {\bibinfo {title} {Swimming
  bacteria power microscopic gears},}\ }\href@noop {} {\bibfield  {journal}
  {\bibinfo  {journal} {PNAS}\ }\textbf {\bibinfo {volume} {107}},\ \bibinfo
  {pages} {969--974} (\bibinfo {year} {2010})}\BibitemShut {NoStop}%
\bibitem [{\citenamefont {Kaiser}\ \emph {et~al.}(2014)\citenamefont {Kaiser},
  \citenamefont {Peshkov}, \citenamefont {Sokolov}, \citenamefont {ten Hagen},
  \citenamefont {L{\"o}wen},\ and\ \citenamefont
  {Aranson}}]{kaiser2014transport}%
  \BibitemOpen
  \bibfield  {author} {\bibinfo {author} {\bibfnamefont {Andreas}\ \bibnamefont
  {Kaiser}}, \bibinfo {author} {\bibfnamefont {Anton}\ \bibnamefont {Peshkov}},
  \bibinfo {author} {\bibfnamefont {Andrey}\ \bibnamefont {Sokolov}}, \bibinfo
  {author} {\bibfnamefont {Borge}\ \bibnamefont {ten Hagen}}, \bibinfo {author}
  {\bibfnamefont {Hartmut}\ \bibnamefont {L{\"o}wen}}, \ and\ \bibinfo {author}
  {\bibfnamefont {Igor~S}\ \bibnamefont {Aranson}},\ }\bibfield  {title}
  {\enquote {\bibinfo {title} {Transport powered by bacterial turbulence},}\
  }\href@noop {} {\bibfield  {journal} {\bibinfo  {journal} {Phys. Rev. Lett.}\
  }\textbf {\bibinfo {volume} {112}},\ \bibinfo {pages} {158101} (\bibinfo
  {year} {2014})}\BibitemShut {NoStop}%
\bibitem [{\citenamefont {Kim}\ and\ \citenamefont
  {Breuer}(2004)}]{kim2004enhanced}%
  \BibitemOpen
  \bibfield  {author} {\bibinfo {author} {\bibfnamefont {Min~Jun}\ \bibnamefont
  {Kim}}\ and\ \bibinfo {author} {\bibfnamefont {Kenneth~S}\ \bibnamefont
  {Breuer}},\ }\bibfield  {title} {\enquote {\bibinfo {title} {Enhanced
  diffusion due to motile bacteria},}\ }\href@noop {} {\bibfield  {journal}
  {\bibinfo  {journal} {Phys. Fluids}\ }\textbf {\bibinfo {volume} {16}},\
  \bibinfo {pages} {L78--L81} (\bibinfo {year} {2004})}\BibitemShut {NoStop}%
\bibitem [{\citenamefont {Underhill}\ \emph {et~al.}(2008)\citenamefont
  {Underhill}, \citenamefont {Hernandez-Ortiz},\ and\ \citenamefont
  {Graham}}]{underhill2008diffusion}%
  \BibitemOpen
  \bibfield  {author} {\bibinfo {author} {\bibfnamefont {Patrick~T}\
  \bibnamefont {Underhill}}, \bibinfo {author} {\bibfnamefont {Juan~P}\
  \bibnamefont {Hernandez-Ortiz}}, \ and\ \bibinfo {author} {\bibfnamefont
  {Michael~D}\ \bibnamefont {Graham}},\ }\bibfield  {title} {\enquote {\bibinfo
  {title} {Diffusion and spatial correlations in suspensions of swimming
  particles},}\ }\href@noop {} {\bibfield  {journal} {\bibinfo  {journal}
  {Phys. Rev. Lett.}\ }\textbf {\bibinfo {volume} {100}},\ \bibinfo {pages}
  {248101} (\bibinfo {year} {2008})}\BibitemShut {NoStop}%
\bibitem [{\citenamefont {Kurtuldu}\ \emph {et~al.}(2011)\citenamefont
  {Kurtuldu}, \citenamefont {Guasto}, \citenamefont {Johnson},\ and\
  \citenamefont {Gollub}}]{kurtuldu2011enhancement}%
  \BibitemOpen
  \bibfield  {author} {\bibinfo {author} {\bibfnamefont {H{\"u}seyin}\
  \bibnamefont {Kurtuldu}}, \bibinfo {author} {\bibfnamefont {Jeffrey~S}\
  \bibnamefont {Guasto}}, \bibinfo {author} {\bibfnamefont {Karl~A}\
  \bibnamefont {Johnson}}, \ and\ \bibinfo {author} {\bibfnamefont {Jerry~P}\
  \bibnamefont {Gollub}},\ }\bibfield  {title} {\enquote {\bibinfo {title}
  {Enhancement of biomixing by swimming algal cells in two-dimensional
  films},}\ }\href@noop {} {\bibfield  {journal} {\bibinfo  {journal} {PNAS}\
  }\textbf {\bibinfo {volume} {108}},\ \bibinfo {pages} {10391--10395}
  (\bibinfo {year} {2011})}\BibitemShut {NoStop}%
\bibitem [{\citenamefont {Jepson}\ \emph {et~al.}(2013)\citenamefont {Jepson},
  \citenamefont {Martinez}, \citenamefont {Schwarz-Linek}, \citenamefont
  {Morozov},\ and\ \citenamefont {Poon}}]{jepson2013enhanced}%
  \BibitemOpen
  \bibfield  {author} {\bibinfo {author} {\bibfnamefont {Alys}\ \bibnamefont
  {Jepson}}, \bibinfo {author} {\bibfnamefont {Vincent~A}\ \bibnamefont
  {Martinez}}, \bibinfo {author} {\bibfnamefont {Jana}\ \bibnamefont
  {Schwarz-Linek}}, \bibinfo {author} {\bibfnamefont {Alexander}\ \bibnamefont
  {Morozov}}, \ and\ \bibinfo {author} {\bibfnamefont {Wilson~CK}\ \bibnamefont
  {Poon}},\ }\bibfield  {title} {\enquote {\bibinfo {title} {Enhanced diffusion
  of nonswimmers in a three-dimensional bath of motile bacteria},}\ }\href@noop
  {} {\bibfield  {journal} {\bibinfo  {journal} {Phys. Rev. E}\ }\textbf
  {\bibinfo {volume} {88}},\ \bibinfo {pages} {041002} (\bibinfo {year}
  {2013})}\BibitemShut {NoStop}%
\bibitem [{\citenamefont {Chen}\ \emph {et~al.}(2007)\citenamefont {Chen},
  \citenamefont {Lau}, \citenamefont {Hough}, \citenamefont {Islam},
  \citenamefont {Goulian}, \citenamefont {Lubensky},\ and\ \citenamefont
  {Yodh}}]{chen2007fluctuations}%
  \BibitemOpen
  \bibfield  {author} {\bibinfo {author} {\bibfnamefont {Daniel~TN}\
  \bibnamefont {Chen}}, \bibinfo {author} {\bibfnamefont {AWC}\ \bibnamefont
  {Lau}}, \bibinfo {author} {\bibfnamefont {Lawrence~A}\ \bibnamefont {Hough}},
  \bibinfo {author} {\bibfnamefont {Mohammad~F}\ \bibnamefont {Islam}},
  \bibinfo {author} {\bibfnamefont {Mark}\ \bibnamefont {Goulian}}, \bibinfo
  {author} {\bibfnamefont {Thomas~C}\ \bibnamefont {Lubensky}}, \ and\ \bibinfo
  {author} {\bibfnamefont {Arjun~G}\ \bibnamefont {Yodh}},\ }\bibfield  {title}
  {\enquote {\bibinfo {title} {Fluctuations and rheology in active bacterial
  suspensions},}\ }\href@noop {} {\bibfield  {journal} {\bibinfo  {journal}
  {Phys. Rev. Lett.}\ }\textbf {\bibinfo {volume} {99}},\ \bibinfo {pages}
  {148302} (\bibinfo {year} {2007})}\BibitemShut {NoStop}%
\bibitem [{\citenamefont {Valeriani}\ \emph {et~al.}(2011)\citenamefont
  {Valeriani}, \citenamefont {Li}, \citenamefont {Novosel}, \citenamefont
  {Arlt},\ and\ \citenamefont {Marenduzzo}}]{valeriani2011colloids}%
  \BibitemOpen
  \bibfield  {author} {\bibinfo {author} {\bibfnamefont {Chantal}\ \bibnamefont
  {Valeriani}}, \bibinfo {author} {\bibfnamefont {Martin}\ \bibnamefont {Li}},
  \bibinfo {author} {\bibfnamefont {John}\ \bibnamefont {Novosel}}, \bibinfo
  {author} {\bibfnamefont {Jochen}\ \bibnamefont {Arlt}}, \ and\ \bibinfo
  {author} {\bibfnamefont {Davide}\ \bibnamefont {Marenduzzo}},\ }\bibfield
  {title} {\enquote {\bibinfo {title} {Colloids in a bacterial bath:
  simulations and experiments},}\ }\href@noop {} {\bibfield  {journal}
  {\bibinfo  {journal} {Soft Matter}\ }\textbf {\bibinfo {volume} {7}},\
  \bibinfo {pages} {5228--5238} (\bibinfo {year} {2011})}\BibitemShut {NoStop}%
\bibitem [{\citenamefont {Mi{\~n}o}\ \emph {et~al.}(2013)\citenamefont
  {Mi{\~n}o}, \citenamefont {Dunstan}, \citenamefont {Rousselet}, \citenamefont
  {Cl{\'e}ment},\ and\ \citenamefont {Soto}}]{mino2013induced}%
  \BibitemOpen
  \bibfield  {author} {\bibinfo {author} {\bibfnamefont {GL}~\bibnamefont
  {Mi{\~n}o}}, \bibinfo {author} {\bibfnamefont {Jocelyn}\ \bibnamefont
  {Dunstan}}, \bibinfo {author} {\bibfnamefont {Annie}\ \bibnamefont
  {Rousselet}}, \bibinfo {author} {\bibfnamefont {E}~\bibnamefont
  {Cl{\'e}ment}}, \ and\ \bibinfo {author} {\bibfnamefont {Rodrigo}\
  \bibnamefont {Soto}},\ }\bibfield  {title} {\enquote {\bibinfo {title}
  {Induced diffusion of tracers in a bacterial suspension: theory and
  experiments},}\ }\href@noop {} {\bibfield  {journal} {\bibinfo  {journal} {J.
  Fluid Mech.}\ }\textbf {\bibinfo {volume} {729}},\ \bibinfo {pages}
  {423--444} (\bibinfo {year} {2013})}\BibitemShut {NoStop}%
\bibitem [{\citenamefont {Ortlieb}\ \emph {et~al.}(2019)\citenamefont
  {Ortlieb}, \citenamefont {Rafa{\"\i}}, \citenamefont {Peyla}, \citenamefont
  {Wagner},\ and\ \citenamefont {John}}]{ortlieb2019statistics}%
  \BibitemOpen
  \bibfield  {author} {\bibinfo {author} {\bibfnamefont {Levke}\ \bibnamefont
  {Ortlieb}}, \bibinfo {author} {\bibfnamefont {Salima}\ \bibnamefont
  {Rafa{\"\i}}}, \bibinfo {author} {\bibfnamefont {Philippe}\ \bibnamefont
  {Peyla}}, \bibinfo {author} {\bibfnamefont {Christian}\ \bibnamefont
  {Wagner}}, \ and\ \bibinfo {author} {\bibfnamefont {Thomas}\ \bibnamefont
  {John}},\ }\bibfield  {title} {\enquote {\bibinfo {title} {Statistics of
  colloidal suspensions stirred by microswimmers},}\ }\href@noop {} {\bibfield
  {journal} {\bibinfo  {journal} {Phys. Rev. Lett.}\ }\textbf {\bibinfo
  {volume} {122}},\ \bibinfo {pages} {148101} (\bibinfo {year}
  {2019})}\BibitemShut {NoStop}%
\bibitem [{\citenamefont {Maggi}\ \emph {et~al.}(2017)\citenamefont {Maggi},
  \citenamefont {Paoluzzi}, \citenamefont {Angelani},\ and\ \citenamefont
  {Di~Leonardo}}]{maggi2017memory}%
  \BibitemOpen
  \bibfield  {author} {\bibinfo {author} {\bibfnamefont {Claudio}\ \bibnamefont
  {Maggi}}, \bibinfo {author} {\bibfnamefont {Matteo}\ \bibnamefont
  {Paoluzzi}}, \bibinfo {author} {\bibfnamefont {Luca}\ \bibnamefont
  {Angelani}}, \ and\ \bibinfo {author} {\bibfnamefont {Roberto}\ \bibnamefont
  {Di~Leonardo}},\ }\bibfield  {title} {\enquote {\bibinfo {title} {Memory-less
  response and violation of the fluctuation-dissipation theorem in colloids
  suspended in an active bath},}\ }\href@noop {} {\bibfield  {journal}
  {\bibinfo  {journal} {Sci. Rep.}\ }\textbf {\bibinfo {volume} {7}},\ \bibinfo
  {pages} {1--7} (\bibinfo {year} {2017})}\BibitemShut {NoStop}%
\bibitem [{\citenamefont {Lagarde}\ \emph {et~al.}(2020)\citenamefont
  {Lagarde}, \citenamefont {Dag{\`e}s}, \citenamefont {Nemoto}, \citenamefont
  {D{\'e}mery}, \citenamefont {Bartolo},\ and\ \citenamefont
  {Gibaud}}]{lagarde2020colloidal}%
  \BibitemOpen
  \bibfield  {author} {\bibinfo {author} {\bibfnamefont {Antoine}\ \bibnamefont
  {Lagarde}}, \bibinfo {author} {\bibfnamefont {No{\'e}mie}\ \bibnamefont
  {Dag{\`e}s}}, \bibinfo {author} {\bibfnamefont {Takahiro}\ \bibnamefont
  {Nemoto}}, \bibinfo {author} {\bibfnamefont {Vincent}\ \bibnamefont
  {D{\'e}mery}}, \bibinfo {author} {\bibfnamefont {Denis}\ \bibnamefont
  {Bartolo}}, \ and\ \bibinfo {author} {\bibfnamefont {Thomas}\ \bibnamefont
  {Gibaud}},\ }\bibfield  {title} {\enquote {\bibinfo {title} {Colloidal
  transport in bacteria suspensions: from bacteria collision to anomalous and
  enhanced diffusion},}\ }\href@noop {} {\bibfield  {journal} {\bibinfo
  {journal} {Soft Matter}\ }\textbf {\bibinfo {volume} {16}},\ \bibinfo {pages}
  {7503--7512} (\bibinfo {year} {2020})}\BibitemShut {NoStop}%
\bibitem [{\citenamefont {Guasto}\ \emph {et~al.}(2010)\citenamefont {Guasto},
  \citenamefont {Johnson},\ and\ \citenamefont
  {Gollub}}]{guasto2010oscillatory}%
  \BibitemOpen
  \bibfield  {author} {\bibinfo {author} {\bibfnamefont {Jeffrey~S}\
  \bibnamefont {Guasto}}, \bibinfo {author} {\bibfnamefont {Karl~A}\
  \bibnamefont {Johnson}}, \ and\ \bibinfo {author} {\bibfnamefont {Jerry~P}\
  \bibnamefont {Gollub}},\ }\bibfield  {title} {\enquote {\bibinfo {title}
  {Oscillatory flows induced by microorganisms swimming in two dimensions},}\
  }\href@noop {} {\bibfield  {journal} {\bibinfo  {journal} {Phys. Rev. Lett.}\
  }\textbf {\bibinfo {volume} {105}},\ \bibinfo {pages} {168102} (\bibinfo
  {year} {2010})}\BibitemShut {NoStop}%
\bibitem [{\citenamefont {Drescher}\ \emph {et~al.}(2010)\citenamefont
  {Drescher}, \citenamefont {Goldstein}, \citenamefont {Michel}, \citenamefont
  {Polin},\ and\ \citenamefont {Tuval}}]{drescher2010direct}%
  \BibitemOpen
  \bibfield  {author} {\bibinfo {author} {\bibfnamefont {Knut}\ \bibnamefont
  {Drescher}}, \bibinfo {author} {\bibfnamefont {Raymond~E}\ \bibnamefont
  {Goldstein}}, \bibinfo {author} {\bibfnamefont {Nicolas}\ \bibnamefont
  {Michel}}, \bibinfo {author} {\bibfnamefont {Marco}\ \bibnamefont {Polin}}, \
  and\ \bibinfo {author} {\bibfnamefont {Idan}\ \bibnamefont {Tuval}},\
  }\bibfield  {title} {\enquote {\bibinfo {title} {Direct measurement of the
  flow field around swimming microorganisms},}\ }\href@noop {} {\bibfield
  {journal} {\bibinfo  {journal} {Phys. Rev. Lett.}\ }\textbf {\bibinfo
  {volume} {105}},\ \bibinfo {pages} {168101} (\bibinfo {year}
  {2010})}\BibitemShut {NoStop}%
\bibitem [{\citenamefont {Leptos}\ \emph {et~al.}(2009)\citenamefont {Leptos},
  \citenamefont {Guasto}, \citenamefont {Gollub}, \citenamefont {Pesci},\ and\
  \citenamefont {Goldstein}}]{leptos2009dynamics}%
  \BibitemOpen
  \bibfield  {author} {\bibinfo {author} {\bibfnamefont {Kyriacos~C}\
  \bibnamefont {Leptos}}, \bibinfo {author} {\bibfnamefont {Jeffrey~S}\
  \bibnamefont {Guasto}}, \bibinfo {author} {\bibfnamefont {Jerry~P}\
  \bibnamefont {Gollub}}, \bibinfo {author} {\bibfnamefont {Adriana~I}\
  \bibnamefont {Pesci}}, \ and\ \bibinfo {author} {\bibfnamefont {Raymond~E}\
  \bibnamefont {Goldstein}},\ }\bibfield  {title} {\enquote {\bibinfo {title}
  {Dynamics of enhanced tracer diffusion in suspensions of swimming eukaryotic
  microorganisms},}\ }\href@noop {} {\bibfield  {journal} {\bibinfo  {journal}
  {Phys. Rev. Lett.}\ }\textbf {\bibinfo {volume} {103}},\ \bibinfo {pages}
  {198103} (\bibinfo {year} {2009})}\BibitemShut {NoStop}%
\bibitem [{\citenamefont {Dunkel}\ \emph {et~al.}(2010)\citenamefont {Dunkel},
  \citenamefont {Putz}, \citenamefont {Zaid},\ and\ \citenamefont
  {Yeomans}}]{dunkel2010swimmer}%
  \BibitemOpen
  \bibfield  {author} {\bibinfo {author} {\bibfnamefont {J{\"o}rn}\
  \bibnamefont {Dunkel}}, \bibinfo {author} {\bibfnamefont {Victor~B}\
  \bibnamefont {Putz}}, \bibinfo {author} {\bibfnamefont {Irwin~M}\
  \bibnamefont {Zaid}}, \ and\ \bibinfo {author} {\bibfnamefont {Julia~M}\
  \bibnamefont {Yeomans}},\ }\bibfield  {title} {\enquote {\bibinfo {title}
  {Swimmer-tracer scattering at low reynolds number},}\ }\href@noop {}
  {\bibfield  {journal} {\bibinfo  {journal} {Soft Matter}\ }\textbf {\bibinfo
  {volume} {6}},\ \bibinfo {pages} {4268--4276} (\bibinfo {year}
  {2010})}\BibitemShut {NoStop}%
\bibitem [{\citenamefont {Klindt}\ and\ \citenamefont
  {Friedrich}(2015)}]{klindt2015flagellar}%
  \BibitemOpen
  \bibfield  {author} {\bibinfo {author} {\bibfnamefont {Gary~S}\ \bibnamefont
  {Klindt}}\ and\ \bibinfo {author} {\bibfnamefont {Benjamin~M}\ \bibnamefont
  {Friedrich}},\ }\bibfield  {title} {\enquote {\bibinfo {title} {Flagellar
  swimmers oscillate between pusher-and puller-type swimming},}\ }\href@noop {}
  {\bibfield  {journal} {\bibinfo  {journal} {Phys. Rev. E}\ }\textbf {\bibinfo
  {volume} {92}},\ \bibinfo {pages} {063019} (\bibinfo {year}
  {2015})}\BibitemShut {NoStop}%
\bibitem [{\citenamefont {Jeanneret}\ \emph {et~al.}(2016)\citenamefont
  {Jeanneret}, \citenamefont {Pushkin}, \citenamefont {Kantsler},\ and\
  \citenamefont {Polin}}]{jeanneret2016entrainment}%
  \BibitemOpen
  \bibfield  {author} {\bibinfo {author} {\bibfnamefont {Rapha{\"e}l}\
  \bibnamefont {Jeanneret}}, \bibinfo {author} {\bibfnamefont {Dmitri~O}\
  \bibnamefont {Pushkin}}, \bibinfo {author} {\bibfnamefont {Vasily}\
  \bibnamefont {Kantsler}}, \ and\ \bibinfo {author} {\bibfnamefont {Marco}\
  \bibnamefont {Polin}},\ }\bibfield  {title} {\enquote {\bibinfo {title}
  {Entrainment dominates the interaction of microalgae with micron-sized
  objects},}\ }\href@noop {} {\bibfield  {journal} {\bibinfo  {journal} {Nat.
  Commun.}\ }\textbf {\bibinfo {volume} {7}},\ \bibinfo {pages} {1--7}
  (\bibinfo {year} {2016})}\BibitemShut {NoStop}%
\bibitem [{\citenamefont {Kasyap}\ \emph {et~al.}(2014)\citenamefont {Kasyap},
  \citenamefont {Koch},\ and\ \citenamefont {Wu}}]{kasyap2014hydrodynamic}%
  \BibitemOpen
  \bibfield  {author} {\bibinfo {author} {\bibfnamefont {TV}~\bibnamefont
  {Kasyap}}, \bibinfo {author} {\bibfnamefont {Donald~L}\ \bibnamefont {Koch}},
  \ and\ \bibinfo {author} {\bibfnamefont {Mingming}\ \bibnamefont {Wu}},\
  }\bibfield  {title} {\enquote {\bibinfo {title} {Hydrodynamic tracer
  diffusion in suspensions of swimming bacteria},}\ }\href@noop {} {\bibfield
  {journal} {\bibinfo  {journal} {Phys. Fluids}\ }\textbf {\bibinfo {volume}
  {26}},\ \bibinfo {pages} {081901} (\bibinfo {year} {2014})}\BibitemShut
  {NoStop}%
\bibitem [{\citenamefont {Morozov}\ and\ \citenamefont
  {Marenduzzo}(2014)}]{morozov2014enhanced}%
  \BibitemOpen
  \bibfield  {author} {\bibinfo {author} {\bibfnamefont {Alexander}\
  \bibnamefont {Morozov}}\ and\ \bibinfo {author} {\bibfnamefont {Davide}\
  \bibnamefont {Marenduzzo}},\ }\bibfield  {title} {\enquote {\bibinfo {title}
  {Enhanced diffusion of tracer particles in dilute bacterial suspensions},}\
  }\href@noop {} {\bibfield  {journal} {\bibinfo  {journal} {Soft Matter}\
  }\textbf {\bibinfo {volume} {10}},\ \bibinfo {pages} {2748--2758} (\bibinfo
  {year} {2014})}\BibitemShut {NoStop}%
\bibitem [{\citenamefont {Burkholder}\ and\ \citenamefont
  {Brady}(2017)}]{burkholder2017tracer}%
  \BibitemOpen
  \bibfield  {author} {\bibinfo {author} {\bibfnamefont {Eric~W}\ \bibnamefont
  {Burkholder}}\ and\ \bibinfo {author} {\bibfnamefont {John~F}\ \bibnamefont
  {Brady}},\ }\bibfield  {title} {\enquote {\bibinfo {title} {Tracer diffusion
  in active suspensions},}\ }\href@noop {} {\bibfield  {journal} {\bibinfo
  {journal} {Phys. Rev. E}\ }\textbf {\bibinfo {volume} {95}},\ \bibinfo
  {pages} {052605} (\bibinfo {year} {2017})}\BibitemShut {NoStop}%
\bibitem [{\citenamefont {Maes}(2020)}]{maes2020fluctuating}%
  \BibitemOpen
  \bibfield  {author} {\bibinfo {author} {\bibfnamefont {Christian}\
  \bibnamefont {Maes}},\ }\bibfield  {title} {\enquote {\bibinfo {title}
  {Fluctuating motion in an active environment},}\ }\href@noop {} {\bibfield
  {journal} {\bibinfo  {journal} {Phys. Rev. Lett.}\ }\textbf {\bibinfo
  {volume} {125}},\ \bibinfo {pages} {208001} (\bibinfo {year}
  {2020})}\BibitemShut {NoStop}%
\bibitem [{\citenamefont {Solon}\ and\ \citenamefont
  {Horowitz}(2022)}]{solon2022einstein}%
  \BibitemOpen
  \bibfield  {author} {\bibinfo {author} {\bibfnamefont {Alexandre}\
  \bibnamefont {Solon}}\ and\ \bibinfo {author} {\bibfnamefont {Jordan~M}\
  \bibnamefont {Horowitz}},\ }\bibfield  {title} {\enquote {\bibinfo {title}
  {On the einstein relation between mobility and diffusion coefficient in an
  active bath},}\ }\href@noop {} {\bibfield  {journal} {\bibinfo  {journal} {J.
  Phys. A Math. Theor.}\ }\textbf {\bibinfo {volume} {55}},\ \bibinfo {pages}
  {184002} (\bibinfo {year} {2022})}\BibitemShut {NoStop}%
\bibitem [{\citenamefont {Kanazawa}\ \emph {et~al.}(2020)\citenamefont
  {Kanazawa}, \citenamefont {Sano}, \citenamefont {Cairoli},\ and\
  \citenamefont {Baule}}]{kanazawa2020loopy}%
  \BibitemOpen
  \bibfield  {author} {\bibinfo {author} {\bibfnamefont {Kiyoshi}\ \bibnamefont
  {Kanazawa}}, \bibinfo {author} {\bibfnamefont {Tomohiko~G}\ \bibnamefont
  {Sano}}, \bibinfo {author} {\bibfnamefont {Andrea}\ \bibnamefont {Cairoli}},
  \ and\ \bibinfo {author} {\bibfnamefont {Adrian}\ \bibnamefont {Baule}},\
  }\bibfield  {title} {\enquote {\bibinfo {title} {Loopy l{\'e}vy flights
  enhance tracer diffusion in active suspensions},}\ }\href@noop {} {\bibfield
  {journal} {\bibinfo  {journal} {Nature}\ }\textbf {\bibinfo {volume} {579}},\
  \bibinfo {pages} {364--367} (\bibinfo {year} {2020})}\BibitemShut {NoStop}%
\bibitem [{\citenamefont {Saintillan}\ and\ \citenamefont
  {Shelley}(2008)}]{saintillan2008instabilities}%
  \BibitemOpen
  \bibfield  {author} {\bibinfo {author} {\bibfnamefont {David}\ \bibnamefont
  {Saintillan}}\ and\ \bibinfo {author} {\bibfnamefont {Michael~J}\
  \bibnamefont {Shelley}},\ }\bibfield  {title} {\enquote {\bibinfo {title}
  {Instabilities and pattern formation in active particle suspensions: kinetic
  theory and continuum simulations},}\ }\href@noop {} {\bibfield  {journal}
  {\bibinfo  {journal} {Phys. Rev. Lett.}\ }\textbf {\bibinfo {volume} {100}},\
  \bibinfo {pages} {178103} (\bibinfo {year} {2008})}\BibitemShut {NoStop}%
\bibitem [{\citenamefont {Lin}\ \emph {et~al.}(2011)\citenamefont {Lin},
  \citenamefont {Thiffeault},\ and\ \citenamefont
  {Childress}}]{lin2011stirring}%
  \BibitemOpen
  \bibfield  {author} {\bibinfo {author} {\bibfnamefont {Zhi}\ \bibnamefont
  {Lin}}, \bibinfo {author} {\bibfnamefont {Jean-Luc}\ \bibnamefont
  {Thiffeault}}, \ and\ \bibinfo {author} {\bibfnamefont {Stephen}\
  \bibnamefont {Childress}},\ }\bibfield  {title} {\enquote {\bibinfo {title}
  {Stirring by squirmers},}\ }\href@noop {} {\bibfield  {journal} {\bibinfo
  {journal} {J. Fluid Mech.}\ }\textbf {\bibinfo {volume} {669}},\ \bibinfo
  {pages} {167--177} (\bibinfo {year} {2011})}\BibitemShut {NoStop}%
\bibitem [{\citenamefont {Thiffeault}(2015)}]{thiffeault2015distribution}%
  \BibitemOpen
  \bibfield  {author} {\bibinfo {author} {\bibfnamefont {Jean-Luc}\
  \bibnamefont {Thiffeault}},\ }\bibfield  {title} {\enquote {\bibinfo {title}
  {Distribution of particle displacements due to swimming microorganisms},}\
  }\href@noop {} {\bibfield  {journal} {\bibinfo  {journal} {Phys. Rev. E}\
  }\textbf {\bibinfo {volume} {92}},\ \bibinfo {pages} {023023} (\bibinfo
  {year} {2015})}\BibitemShut {NoStop}%
\bibitem [{\citenamefont {Brown}\ \emph {et~al.}(2016)\citenamefont {Brown},
  \citenamefont {Vladescu}, \citenamefont {Dawson}, \citenamefont {Vissers},
  \citenamefont {Schwarz-Linek}, \citenamefont {Lintuvuori},\ and\
  \citenamefont {Poon}}]{brown2016swimming}%
  \BibitemOpen
  \bibfield  {author} {\bibinfo {author} {\bibfnamefont {Aidan~T}\ \bibnamefont
  {Brown}}, \bibinfo {author} {\bibfnamefont {Ioana~D}\ \bibnamefont
  {Vladescu}}, \bibinfo {author} {\bibfnamefont {Angela}\ \bibnamefont
  {Dawson}}, \bibinfo {author} {\bibfnamefont {Teun}\ \bibnamefont {Vissers}},
  \bibinfo {author} {\bibfnamefont {Jana}\ \bibnamefont {Schwarz-Linek}},
  \bibinfo {author} {\bibfnamefont {Juho~S}\ \bibnamefont {Lintuvuori}}, \ and\
  \bibinfo {author} {\bibfnamefont {Wilson~CK}\ \bibnamefont {Poon}},\
  }\bibfield  {title} {\enquote {\bibinfo {title} {Swimming in a crystal},}\
  }\href@noop {} {\bibfield  {journal} {\bibinfo  {journal} {Soft Matter}\
  }\textbf {\bibinfo {volume} {12}},\ \bibinfo {pages} {131--140} (\bibinfo
  {year} {2016})}\BibitemShut {NoStop}%
\bibitem [{\citenamefont {K{\"u}mmel}\ \emph {et~al.}(2015)\citenamefont
  {K{\"u}mmel}, \citenamefont {Shabestari}, \citenamefont {Lozano},
  \citenamefont {Volpe},\ and\ \citenamefont
  {Bechinger}}]{kummel2015formation}%
  \BibitemOpen
  \bibfield  {author} {\bibinfo {author} {\bibfnamefont {Felix}\ \bibnamefont
  {K{\"u}mmel}}, \bibinfo {author} {\bibfnamefont {Parmida}\ \bibnamefont
  {Shabestari}}, \bibinfo {author} {\bibfnamefont {Celia}\ \bibnamefont
  {Lozano}}, \bibinfo {author} {\bibfnamefont {Giovanni}\ \bibnamefont
  {Volpe}}, \ and\ \bibinfo {author} {\bibfnamefont {Clemens}\ \bibnamefont
  {Bechinger}},\ }\bibfield  {title} {\enquote {\bibinfo {title} {Formation,
  compression and surface melting of colloidal clusters by active particles},}\
  }\href@noop {} {\bibfield  {journal} {\bibinfo  {journal} {Soft Matter}\
  }\textbf {\bibinfo {volume} {11}},\ \bibinfo {pages} {6187--6191} (\bibinfo
  {year} {2015})}\BibitemShut {NoStop}%
\bibitem [{\citenamefont {Ramananarivo}\ \emph {et~al.}(2019)\citenamefont
  {Ramananarivo}, \citenamefont {Ducrot},\ and\ \citenamefont
  {Palacci}}]{ramananarivo2019activity}%
  \BibitemOpen
  \bibfield  {author} {\bibinfo {author} {\bibfnamefont {Sophie}\ \bibnamefont
  {Ramananarivo}}, \bibinfo {author} {\bibfnamefont {Etienne}\ \bibnamefont
  {Ducrot}}, \ and\ \bibinfo {author} {\bibfnamefont {Jeremie}\ \bibnamefont
  {Palacci}},\ }\bibfield  {title} {\enquote {\bibinfo {title}
  {Activity-controlled annealing of colloidal monolayers},}\ }\href@noop {}
  {\bibfield  {journal} {\bibinfo  {journal} {Nat. Commun}\ }\textbf {\bibinfo
  {volume} {10}},\ \bibinfo {pages} {3380} (\bibinfo {year}
  {2019})}\BibitemShut {NoStop}%
\bibitem [{\citenamefont {Jakuszeit}\ \emph {et~al.}(2019)\citenamefont
  {Jakuszeit}, \citenamefont {Croze},\ and\ \citenamefont
  {Bell}}]{jakuszeit2019diffusion}%
  \BibitemOpen
  \bibfield  {author} {\bibinfo {author} {\bibfnamefont {Theresa}\ \bibnamefont
  {Jakuszeit}}, \bibinfo {author} {\bibfnamefont {Ottavio~A}\ \bibnamefont
  {Croze}}, \ and\ \bibinfo {author} {\bibfnamefont {Samuel}\ \bibnamefont
  {Bell}},\ }\bibfield  {title} {\enquote {\bibinfo {title} {Diffusion of
  active particles in a complex environment: Role of surface scattering},}\
  }\href@noop {} {\bibfield  {journal} {\bibinfo  {journal} {Phys. Rev. E}\
  }\textbf {\bibinfo {volume} {99}},\ \bibinfo {pages} {012610} (\bibinfo
  {year} {2019})}\BibitemShut {NoStop}%
\bibitem [{\citenamefont {Saintillan}(2023)}]{saintillan2023dispersion}%
  \BibitemOpen
  \bibfield  {author} {\bibinfo {author} {\bibfnamefont {David}\ \bibnamefont
  {Saintillan}},\ }\bibfield  {title} {\enquote {\bibinfo {title} {Dispersion
  of run-and-tumble microswimmers through disordered media},}\ }\href@noop {}
  {\bibfield  {journal} {\bibinfo  {journal} {Phys. Rev. E}\ }\textbf {\bibinfo
  {volume} {108}},\ \bibinfo {pages} {064608} (\bibinfo {year}
  {2023})}\BibitemShut {NoStop}%
\bibitem [{\citenamefont {Patteson}\ \emph {et~al.}(2016)\citenamefont
  {Patteson}, \citenamefont {Gopinath}, \citenamefont {Purohit},\ and\
  \citenamefont {Arratia}}]{patteson2016particle}%
  \BibitemOpen
  \bibfield  {author} {\bibinfo {author} {\bibfnamefont {Alison~E}\
  \bibnamefont {Patteson}}, \bibinfo {author} {\bibfnamefont {Arvind}\
  \bibnamefont {Gopinath}}, \bibinfo {author} {\bibfnamefont {Prashant~K}\
  \bibnamefont {Purohit}}, \ and\ \bibinfo {author} {\bibfnamefont {Paulo~E}\
  \bibnamefont {Arratia}},\ }\bibfield  {title} {\enquote {\bibinfo {title}
  {Particle diffusion in active fluids is non-monotonic in size},}\ }\href@noop
  {} {\bibfield  {journal} {\bibinfo  {journal} {Soft Matter}\ }\textbf
  {\bibinfo {volume} {12}},\ \bibinfo {pages} {2365--2372} (\bibinfo {year}
  {2016})}\BibitemShut {NoStop}%
\bibitem [{\citenamefont {Berg}\ and\ \citenamefont
  {Purcell}(1977)}]{berg1977physics}%
  \BibitemOpen
  \bibfield  {author} {\bibinfo {author} {\bibfnamefont {Howard~C}\
  \bibnamefont {Berg}}\ and\ \bibinfo {author} {\bibfnamefont {Edward~M}\
  \bibnamefont {Purcell}},\ }\bibfield  {title} {\enquote {\bibinfo {title}
  {Physics of chemoreception},}\ }\href@noop {} {\bibfield  {journal} {\bibinfo
   {journal} {Biophys. J.}\ }\textbf {\bibinfo {volume} {20}},\ \bibinfo
  {pages} {193--219} (\bibinfo {year} {1977})}\BibitemShut {NoStop}%
\bibitem [{\citenamefont {Berg}(2018)}]{berg2018random}%
  \BibitemOpen
  \bibfield  {author} {\bibinfo {author} {\bibfnamefont {Howard~C}\
  \bibnamefont {Berg}},\ }\bibfield  {title} {\enquote {\bibinfo {title}
  {Random walks in biology},}\ }in\ \href@noop {} {\emph {\bibinfo {booktitle}
  {Random Walks in Biology}}}\ (\bibinfo  {publisher} {Princeton University
  Press},\ \bibinfo {year} {2018})\BibitemShut {NoStop}%
\bibitem [{\citenamefont {Block}\ \emph {et~al.}(1983)\citenamefont {Block},
  \citenamefont {Segall},\ and\ \citenamefont {Berg}}]{block1983adaptation}%
  \BibitemOpen
  \bibfield  {author} {\bibinfo {author} {\bibfnamefont {Steven~M}\
  \bibnamefont {Block}}, \bibinfo {author} {\bibfnamefont {Jeffery~E}\
  \bibnamefont {Segall}}, \ and\ \bibinfo {author} {\bibfnamefont {Howard~C}\
  \bibnamefont {Berg}},\ }\bibfield  {title} {\enquote {\bibinfo {title}
  {Adaptation kinetics in bacterial chemotaxis},}\ }\href@noop {} {\bibfield
  {journal} {\bibinfo  {journal} {J. Bacteriol.}\ }\textbf {\bibinfo {volume}
  {154}},\ \bibinfo {pages} {312--323} (\bibinfo {year} {1983})}\BibitemShut
  {NoStop}%
\bibitem [{\citenamefont {Alon}\ \emph {et~al.}(1998)\citenamefont {Alon},
  \citenamefont {Camarena}, \citenamefont {Surette}, \citenamefont {y~Arcas},
  \citenamefont {Liu}, \citenamefont {Leibler},\ and\ \citenamefont
  {Stock}}]{alon1998response}%
  \BibitemOpen
  \bibfield  {author} {\bibinfo {author} {\bibfnamefont {Uri}\ \bibnamefont
  {Alon}}, \bibinfo {author} {\bibfnamefont {Laura}\ \bibnamefont {Camarena}},
  \bibinfo {author} {\bibfnamefont {Michael~G}\ \bibnamefont {Surette}},
  \bibinfo {author} {\bibfnamefont {Blaise~Aguera}\ \bibnamefont {y~Arcas}},
  \bibinfo {author} {\bibfnamefont {Yi}~\bibnamefont {Liu}}, \bibinfo {author}
  {\bibfnamefont {Stanislas}\ \bibnamefont {Leibler}}, \ and\ \bibinfo {author}
  {\bibfnamefont {Jeffry~B}\ \bibnamefont {Stock}},\ }\bibfield  {title}
  {\enquote {\bibinfo {title} {Response regulator output in bacterial
  chemotaxis},}\ }\href@noop {} {\bibfield  {journal} {\bibinfo  {journal} {The
  EMBO journal}\ }\textbf {\bibinfo {volume} {17}},\ \bibinfo {pages}
  {4238--4248} (\bibinfo {year} {1998})}\BibitemShut {NoStop}%
\bibitem [{\citenamefont {Saragosti}\ \emph {et~al.}(2011)\citenamefont
  {Saragosti}, \citenamefont {Calvez}, \citenamefont {Bournaveas},
  \citenamefont {Perthame}, \citenamefont {Buguin},\ and\ \citenamefont
  {Silberzan}}]{saragosti2011directional}%
  \BibitemOpen
  \bibfield  {author} {\bibinfo {author} {\bibfnamefont {Jonathan}\
  \bibnamefont {Saragosti}}, \bibinfo {author} {\bibfnamefont {Vincent}\
  \bibnamefont {Calvez}}, \bibinfo {author} {\bibfnamefont {Nikolaos}\
  \bibnamefont {Bournaveas}}, \bibinfo {author} {\bibfnamefont {Beno{\i}t}\
  \bibnamefont {Perthame}}, \bibinfo {author} {\bibfnamefont {Axel}\
  \bibnamefont {Buguin}}, \ and\ \bibinfo {author} {\bibfnamefont {Pascal}\
  \bibnamefont {Silberzan}},\ }\bibfield  {title} {\enquote {\bibinfo {title}
  {Directional persistence of chemotactic bacteria in a traveling concentration
  wave},}\ }\href@noop {} {\bibfield  {journal} {\bibinfo  {journal} {PNAS}\
  }\textbf {\bibinfo {volume} {108}},\ \bibinfo {pages} {16235--16240}
  (\bibinfo {year} {2011})}\BibitemShut {NoStop}%
\bibitem [{\citenamefont {Soto}\ and\ \citenamefont
  {Golestanian}(2014)}]{soto2014run}%
  \BibitemOpen
  \bibfield  {author} {\bibinfo {author} {\bibfnamefont {Rodrigo}\ \bibnamefont
  {Soto}}\ and\ \bibinfo {author} {\bibfnamefont {Ramin}\ \bibnamefont
  {Golestanian}},\ }\bibfield  {title} {\enquote {\bibinfo {title}
  {Run-and-tumble dynamics in a crowded environment: Persistent exclusion
  process for swimmers},}\ }\href@noop {} {\bibfield  {journal} {\bibinfo
  {journal} {Phys. Rev. E}\ }\textbf {\bibinfo {volume} {89}},\ \bibinfo
  {pages} {012706} (\bibinfo {year} {2014})}\BibitemShut {NoStop}%
\bibitem [{\citenamefont {Elgeti}\ and\ \citenamefont
  {Gompper}(2015)}]{elgeti2015run}%
  \BibitemOpen
  \bibfield  {author} {\bibinfo {author} {\bibfnamefont {Jens}\ \bibnamefont
  {Elgeti}}\ and\ \bibinfo {author} {\bibfnamefont {Gerhard}\ \bibnamefont
  {Gompper}},\ }\bibfield  {title} {\enquote {\bibinfo {title} {Run-and-tumble
  dynamics of self-propelled particles in confinement},}\ }\href@noop {}
  {\bibfield  {journal} {\bibinfo  {journal} {EPL}\ }\textbf {\bibinfo {volume}
  {109}},\ \bibinfo {pages} {58003} (\bibinfo {year} {2015})}\BibitemShut
  {NoStop}%
\bibitem [{\citenamefont {Drescher}\ \emph {et~al.}(2011)\citenamefont
  {Drescher}, \citenamefont {Dunkel}, \citenamefont {Cisneros}, \citenamefont
  {Ganguly},\ and\ \citenamefont {Goldstein}}]{drescher2011fluid}%
  \BibitemOpen
  \bibfield  {author} {\bibinfo {author} {\bibfnamefont {Knut}\ \bibnamefont
  {Drescher}}, \bibinfo {author} {\bibfnamefont {J{\"o}rn}\ \bibnamefont
  {Dunkel}}, \bibinfo {author} {\bibfnamefont {Luis~H}\ \bibnamefont
  {Cisneros}}, \bibinfo {author} {\bibfnamefont {Sujoy}\ \bibnamefont
  {Ganguly}}, \ and\ \bibinfo {author} {\bibfnamefont {Raymond~E}\ \bibnamefont
  {Goldstein}},\ }\bibfield  {title} {\enquote {\bibinfo {title} {Fluid
  dynamics and noise in bacterial cell--cell and cell--surface scattering},}\
  }\href@noop {} {\bibfield  {journal} {\bibinfo  {journal} {PNAS}\ }\textbf
  {\bibinfo {volume} {108}},\ \bibinfo {pages} {10940--10945} (\bibinfo {year}
  {2011})}\BibitemShut {NoStop}%
\bibitem [{\citenamefont {Stenhammar}\ \emph {et~al.}(2017)\citenamefont
  {Stenhammar}, \citenamefont {Nardini}, \citenamefont {Nash}, \citenamefont
  {Marenduzzo},\ and\ \citenamefont {Morozov}}]{stenhammar2017role}%
  \BibitemOpen
  \bibfield  {author} {\bibinfo {author} {\bibfnamefont {Joakim}\ \bibnamefont
  {Stenhammar}}, \bibinfo {author} {\bibfnamefont {Cesare}\ \bibnamefont
  {Nardini}}, \bibinfo {author} {\bibfnamefont {Rupert~W}\ \bibnamefont
  {Nash}}, \bibinfo {author} {\bibfnamefont {Davide}\ \bibnamefont
  {Marenduzzo}}, \ and\ \bibinfo {author} {\bibfnamefont {Alexander}\
  \bibnamefont {Morozov}},\ }\bibfield  {title} {\enquote {\bibinfo {title}
  {Role of correlations in the collective behavior of microswimmer
  suspensions},}\ }\href@noop {} {\bibfield  {journal} {\bibinfo  {journal}
  {Phys. Rev. Lett.}\ }\textbf {\bibinfo {volume} {119}},\ \bibinfo {pages}
  {028005} (\bibinfo {year} {2017})}\BibitemShut {NoStop}%
\bibitem [{\citenamefont {Yoshinaga}\ and\ \citenamefont
  {Liverpool}(2018)}]{yoshinaga2018hydrodynamic}%
  \BibitemOpen
  \bibfield  {author} {\bibinfo {author} {\bibfnamefont {Natsuhiko}\
  \bibnamefont {Yoshinaga}}\ and\ \bibinfo {author} {\bibfnamefont
  {Tanniemola~B}\ \bibnamefont {Liverpool}},\ }\bibfield  {title} {\enquote
  {\bibinfo {title} {From hydrodynamic lubrication to many-body interactions in
  dense suspensions of active swimmers},}\ }\href@noop {} {\bibfield  {journal}
  {\bibinfo  {journal} {EPJE}\ }\textbf {\bibinfo {volume} {41}},\ \bibinfo
  {pages} {1--22} (\bibinfo {year} {2018})}\BibitemShut {NoStop}%
\bibitem [{\citenamefont {Berke}\ \emph {et~al.}(2008)\citenamefont {Berke},
  \citenamefont {Turner}, \citenamefont {Berg},\ and\ \citenamefont
  {Lauga}}]{berke2008hydrodynamic}%
  \BibitemOpen
  \bibfield  {author} {\bibinfo {author} {\bibfnamefont {Allison~P}\
  \bibnamefont {Berke}}, \bibinfo {author} {\bibfnamefont {Linda}\ \bibnamefont
  {Turner}}, \bibinfo {author} {\bibfnamefont {Howard~C}\ \bibnamefont {Berg}},
  \ and\ \bibinfo {author} {\bibfnamefont {Eric}\ \bibnamefont {Lauga}},\
  }\bibfield  {title} {\enquote {\bibinfo {title} {Hydrodynamic attraction of
  swimming microorganisms by surfaces},}\ }\href@noop {} {\bibfield  {journal}
  {\bibinfo  {journal} {Phys. Rev. Lett.}\ }\textbf {\bibinfo {volume} {101}},\
  \bibinfo {pages} {038102} (\bibinfo {year} {2008})}\BibitemShut {NoStop}%
\bibitem [{\citenamefont {Lauga}\ \emph {et~al.}(2006)\citenamefont {Lauga},
  \citenamefont {DiLuzio}, \citenamefont {Whitesides},\ and\ \citenamefont
  {Stone}}]{lauga2006swimming}%
  \BibitemOpen
  \bibfield  {author} {\bibinfo {author} {\bibfnamefont {Eric}\ \bibnamefont
  {Lauga}}, \bibinfo {author} {\bibfnamefont {Willow~R}\ \bibnamefont
  {DiLuzio}}, \bibinfo {author} {\bibfnamefont {George~M}\ \bibnamefont
  {Whitesides}}, \ and\ \bibinfo {author} {\bibfnamefont {Howard~A}\
  \bibnamefont {Stone}},\ }\bibfield  {title} {\enquote {\bibinfo {title}
  {Swimming in circles: motion of bacteria near solid boundaries},}\
  }\href@noop {} {\bibfield  {journal} {\bibinfo  {journal} {Biophys. J.}\
  }\textbf {\bibinfo {volume} {90}},\ \bibinfo {pages} {400--412} (\bibinfo
  {year} {2006})}\BibitemShut {NoStop}%
\bibitem [{\citenamefont {Spagnolie}\ \emph {et~al.}(2015)\citenamefont
  {Spagnolie}, \citenamefont {Moreno-Flores}, \citenamefont {Bartolo},\ and\
  \citenamefont {Lauga}}]{spagnolie2015geometric}%
  \BibitemOpen
  \bibfield  {author} {\bibinfo {author} {\bibfnamefont {Saverio~E}\
  \bibnamefont {Spagnolie}}, \bibinfo {author} {\bibfnamefont {Gregorio~R}\
  \bibnamefont {Moreno-Flores}}, \bibinfo {author} {\bibfnamefont {Denis}\
  \bibnamefont {Bartolo}}, \ and\ \bibinfo {author} {\bibfnamefont {Eric}\
  \bibnamefont {Lauga}},\ }\bibfield  {title} {\enquote {\bibinfo {title}
  {Geometric capture and escape of a microswimmer colliding with an
  obstacle},}\ }\href@noop {} {\bibfield  {journal} {\bibinfo  {journal} {Soft
  Matter}\ }\textbf {\bibinfo {volume} {11}},\ \bibinfo {pages} {3396--3411}
  (\bibinfo {year} {2015})}\BibitemShut {NoStop}%
\bibitem [{\citenamefont {Molaei}\ \emph {et~al.}(2014)\citenamefont {Molaei},
  \citenamefont {Barry}, \citenamefont {Stocker},\ and\ \citenamefont
  {Sheng}}]{molaei2014failed}%
  \BibitemOpen
  \bibfield  {author} {\bibinfo {author} {\bibfnamefont {Mehdi}\ \bibnamefont
  {Molaei}}, \bibinfo {author} {\bibfnamefont {Michael}\ \bibnamefont {Barry}},
  \bibinfo {author} {\bibfnamefont {Roman}\ \bibnamefont {Stocker}}, \ and\
  \bibinfo {author} {\bibfnamefont {Jian}\ \bibnamefont {Sheng}},\ }\bibfield
  {title} {\enquote {\bibinfo {title} {Failed escape: solid surfaces prevent
  tumbling of escherichia coli},}\ }\href@noop {} {\bibfield  {journal}
  {\bibinfo  {journal} {Phys. Rev. Lett.}\ }\textbf {\bibinfo {volume} {113}},\
  \bibinfo {pages} {068103} (\bibinfo {year} {2014})}\BibitemShut {NoStop}%
\bibitem [{\citenamefont {Perez~Ipi{\~n}a}\ \emph {et~al.}(2019)\citenamefont
  {Perez~Ipi{\~n}a}, \citenamefont {Otte}, \citenamefont {Pontier-Bres},
  \citenamefont {Czerucka},\ and\ \citenamefont {Peruani}}]{perez2019bacteria}%
  \BibitemOpen
  \bibfield  {author} {\bibinfo {author} {\bibfnamefont {Emiliano}\
  \bibnamefont {Perez~Ipi{\~n}a}}, \bibinfo {author} {\bibfnamefont {Stefan}\
  \bibnamefont {Otte}}, \bibinfo {author} {\bibfnamefont {Rodolphe}\
  \bibnamefont {Pontier-Bres}}, \bibinfo {author} {\bibfnamefont {Dorota}\
  \bibnamefont {Czerucka}}, \ and\ \bibinfo {author} {\bibfnamefont {Fernando}\
  \bibnamefont {Peruani}},\ }\bibfield  {title} {\enquote {\bibinfo {title}
  {Bacteria display optimal transport near surfaces},}\ }\href@noop {}
  {\bibfield  {journal} {\bibinfo  {journal} {Nat. Phys.}\ }\textbf {\bibinfo
  {volume} {15}},\ \bibinfo {pages} {610--615} (\bibinfo {year}
  {2019})}\BibitemShut {NoStop}%
\bibitem [{\citenamefont {Sipos}\ \emph {et~al.}(2015)\citenamefont {Sipos},
  \citenamefont {Nagy}, \citenamefont {Di~Leonardo},\ and\ \citenamefont
  {Galajda}}]{sipos2015hydrodynamic}%
  \BibitemOpen
  \bibfield  {author} {\bibinfo {author} {\bibfnamefont {Orsolya}\ \bibnamefont
  {Sipos}}, \bibinfo {author} {\bibfnamefont {K}~\bibnamefont {Nagy}}, \bibinfo
  {author} {\bibfnamefont {R}~\bibnamefont {Di~Leonardo}}, \ and\ \bibinfo
  {author} {\bibfnamefont {P}~\bibnamefont {Galajda}},\ }\bibfield  {title}
  {\enquote {\bibinfo {title} {Hydrodynamic trapping of swimming bacteria by
  convex walls},}\ }\href@noop {} {\bibfield  {journal} {\bibinfo  {journal}
  {Phys. Rev. Lett.}\ }\textbf {\bibinfo {volume} {114}},\ \bibinfo {pages}
  {258104} (\bibinfo {year} {2015})}\BibitemShut {NoStop}%
\bibitem [{\citenamefont {Takatori}\ \emph {et~al.}(2014)\citenamefont
  {Takatori}, \citenamefont {Yan},\ and\ \citenamefont
  {Brady}}]{takatori2014swim}%
  \BibitemOpen
  \bibfield  {author} {\bibinfo {author} {\bibfnamefont {Sho~C}\ \bibnamefont
  {Takatori}}, \bibinfo {author} {\bibfnamefont {Wen}\ \bibnamefont {Yan}}, \
  and\ \bibinfo {author} {\bibfnamefont {John~F}\ \bibnamefont {Brady}},\
  }\bibfield  {title} {\enquote {\bibinfo {title} {Swim pressure: stress
  generation in active matter},}\ }\href@noop {} {\bibfield  {journal}
  {\bibinfo  {journal} {Phys. Rev. Lett.}\ }\textbf {\bibinfo {volume} {113}},\
  \bibinfo {pages} {028103} (\bibinfo {year} {2014})}\BibitemShut {NoStop}%
\bibitem [{\citenamefont {Takagi}\ \emph {et~al.}(2014)\citenamefont {Takagi},
  \citenamefont {Palacci}, \citenamefont {Braunschweig}, \citenamefont
  {Shelley},\ and\ \citenamefont {Zhang}}]{takagi2014hydrodynamic}%
  \BibitemOpen
  \bibfield  {author} {\bibinfo {author} {\bibfnamefont {Daisuke}\ \bibnamefont
  {Takagi}}, \bibinfo {author} {\bibfnamefont {J{\'e}r{\'e}mie}\ \bibnamefont
  {Palacci}}, \bibinfo {author} {\bibfnamefont {Adam~B}\ \bibnamefont
  {Braunschweig}}, \bibinfo {author} {\bibfnamefont {Michael~J}\ \bibnamefont
  {Shelley}}, \ and\ \bibinfo {author} {\bibfnamefont {Jun}\ \bibnamefont
  {Zhang}},\ }\bibfield  {title} {\enquote {\bibinfo {title} {Hydrodynamic
  capture of microswimmers into sphere-bound orbits},}\ }\href@noop {}
  {\bibfield  {journal} {\bibinfo  {journal} {Soft Matter}\ }\textbf {\bibinfo
  {volume} {10}},\ \bibinfo {pages} {1784--1789} (\bibinfo {year}
  {2014})}\BibitemShut {NoStop}%
\bibitem [{\citenamefont {Omar}\ \emph {et~al.}(2020)\citenamefont {Omar},
  \citenamefont {Wang},\ and\ \citenamefont {Brady}}]{omar2020microscopic}%
  \BibitemOpen
  \bibfield  {author} {\bibinfo {author} {\bibfnamefont {Ahmad~K}\ \bibnamefont
  {Omar}}, \bibinfo {author} {\bibfnamefont {Zhen-Gang}\ \bibnamefont {Wang}},
  \ and\ \bibinfo {author} {\bibfnamefont {John~F}\ \bibnamefont {Brady}},\
  }\bibfield  {title} {\enquote {\bibinfo {title} {Microscopic origins of the
  swim pressure and the anomalous surface tension of active matter},}\
  }\href@noop {} {\bibfield  {journal} {\bibinfo  {journal} {Phys. Rev. E}\
  }\textbf {\bibinfo {volume} {101}},\ \bibinfo {pages} {012604} (\bibinfo
  {year} {2020})}\BibitemShut {NoStop}%
\bibitem [{\citenamefont {Hinch}(1975)}]{hinch1975application}%
  \BibitemOpen
  \bibfield  {author} {\bibinfo {author} {\bibfnamefont {Edward~John}\
  \bibnamefont {Hinch}},\ }\bibfield  {title} {\enquote {\bibinfo {title}
  {Application of the langevin equation to fluid suspensions},}\ }\href@noop {}
  {\bibfield  {journal} {\bibinfo  {journal} {J. Fluid Mech.}\ }\textbf
  {\bibinfo {volume} {72}},\ \bibinfo {pages} {499--511} (\bibinfo {year}
  {1975})}\BibitemShut {NoStop}%
\bibitem [{\citenamefont {Kubo}\ \emph {et~al.}(2012)\citenamefont {Kubo},
  \citenamefont {Toda},\ and\ \citenamefont
  {Hashitsume}}]{kubo2012statistical}%
  \BibitemOpen
  \bibfield  {author} {\bibinfo {author} {\bibfnamefont {Ryogo}\ \bibnamefont
  {Kubo}}, \bibinfo {author} {\bibfnamefont {Morikazu}\ \bibnamefont {Toda}}, \
  and\ \bibinfo {author} {\bibfnamefont {Natsuki}\ \bibnamefont {Hashitsume}},\
  }\href@noop {} {\emph {\bibinfo {title} {Statistical physics II:
  nonequilibrium statistical mechanics}}},\ Vol.~\bibinfo {volume} {31}\
  (\bibinfo  {publisher} {Springer Science \& Business Media},\ \bibinfo {year}
  {2012})\BibitemShut {NoStop}%
\bibitem [{\citenamefont {Chattopadhyay}\ \emph {et~al.}(2006)\citenamefont
  {Chattopadhyay}, \citenamefont {Moldovan}, \citenamefont {Yeung},\ and\
  \citenamefont {Wu}}]{chattopadhyay2006swimming}%
  \BibitemOpen
  \bibfield  {author} {\bibinfo {author} {\bibfnamefont {Suddhashil}\
  \bibnamefont {Chattopadhyay}}, \bibinfo {author} {\bibfnamefont {Radu}\
  \bibnamefont {Moldovan}}, \bibinfo {author} {\bibfnamefont {Chuck}\
  \bibnamefont {Yeung}}, \ and\ \bibinfo {author} {\bibfnamefont
  {XL}~\bibnamefont {Wu}},\ }\bibfield  {title} {\enquote {\bibinfo {title}
  {Swimming efficiency of bacterium escherichia coli},}\ }\href@noop {}
  {\bibfield  {journal} {\bibinfo  {journal} {PNAS}\ }\textbf {\bibinfo
  {volume} {103}},\ \bibinfo {pages} {13712--13717} (\bibinfo {year}
  {2006})}\BibitemShut {NoStop}%
\bibitem [{\citenamefont {Angelani}\ \emph {et~al.}(2011)\citenamefont
  {Angelani}, \citenamefont {Maggi}, \citenamefont {Bernardini}, \citenamefont
  {Rizzo},\ and\ \citenamefont {Di~Leonardo}}]{angelani2011effective}%
  \BibitemOpen
  \bibfield  {author} {\bibinfo {author} {\bibfnamefont {L}~\bibnamefont
  {Angelani}}, \bibinfo {author} {\bibfnamefont {C}~\bibnamefont {Maggi}},
  \bibinfo {author} {\bibfnamefont {ML}~\bibnamefont {Bernardini}}, \bibinfo
  {author} {\bibfnamefont {A}~\bibnamefont {Rizzo}}, \ and\ \bibinfo {author}
  {\bibfnamefont {R}~\bibnamefont {Di~Leonardo}},\ }\bibfield  {title}
  {\enquote {\bibinfo {title} {Effective interactions between colloidal
  particles suspended in a bath of swimming cells},}\ }\href@noop {} {\bibfield
   {journal} {\bibinfo  {journal} {Phys. Rev. Lett.}\ }\textbf {\bibinfo
  {volume} {107}},\ \bibinfo {pages} {138302} (\bibinfo {year}
  {2011})}\BibitemShut {NoStop}%
\bibitem [{\citenamefont {Volpe}\ \emph {et~al.}(2011)\citenamefont {Volpe},
  \citenamefont {Buttinoni}, \citenamefont {Vogt}, \citenamefont
  {K{\"u}mmerer},\ and\ \citenamefont {Bechinger}}]{volpe2011microswimmers}%
  \BibitemOpen
  \bibfield  {author} {\bibinfo {author} {\bibfnamefont {Giovanni}\
  \bibnamefont {Volpe}}, \bibinfo {author} {\bibfnamefont {Ivo}\ \bibnamefont
  {Buttinoni}}, \bibinfo {author} {\bibfnamefont {Dominik}\ \bibnamefont
  {Vogt}}, \bibinfo {author} {\bibfnamefont {Hans-J{\"u}rgen}\ \bibnamefont
  {K{\"u}mmerer}}, \ and\ \bibinfo {author} {\bibfnamefont {Clemens}\
  \bibnamefont {Bechinger}},\ }\bibfield  {title} {\enquote {\bibinfo {title}
  {Microswimmers in patterned environments},}\ }\href@noop {} {\bibfield
  {journal} {\bibinfo  {journal} {Soft Matter}\ }\textbf {\bibinfo {volume}
  {7}},\ \bibinfo {pages} {8810--8815} (\bibinfo {year} {2011})}\BibitemShut
  {NoStop}%
\bibitem [{\citenamefont {Kantsler}\ \emph {et~al.}(2013)\citenamefont
  {Kantsler}, \citenamefont {Dunkel}, \citenamefont {Polin},\ and\
  \citenamefont {Goldstein}}]{kantsler2013ciliary}%
  \BibitemOpen
  \bibfield  {author} {\bibinfo {author} {\bibfnamefont {Vasily}\ \bibnamefont
  {Kantsler}}, \bibinfo {author} {\bibfnamefont {J{\"o}rn}\ \bibnamefont
  {Dunkel}}, \bibinfo {author} {\bibfnamefont {Marco}\ \bibnamefont {Polin}}, \
  and\ \bibinfo {author} {\bibfnamefont {Raymond~E}\ \bibnamefont
  {Goldstein}},\ }\bibfield  {title} {\enquote {\bibinfo {title} {Ciliary
  contact interactions dominate surface scattering of swimming eukaryotes},}\
  }\href@noop {} {\bibfield  {journal} {\bibinfo  {journal} {PNAS}\ }\textbf
  {\bibinfo {volume} {110}},\ \bibinfo {pages} {1187--1192} (\bibinfo {year}
  {2013})}\BibitemShut {NoStop}%
\bibitem [{\citenamefont {Korobkova}\ \emph {et~al.}(2004)\citenamefont
  {Korobkova}, \citenamefont {Emonet}, \citenamefont {Vilar}, \citenamefont
  {Shimizu},\ and\ \citenamefont {Cluzel}}]{korobkova2004molecular}%
  \BibitemOpen
  \bibfield  {author} {\bibinfo {author} {\bibfnamefont {Ekaterina}\
  \bibnamefont {Korobkova}}, \bibinfo {author} {\bibfnamefont {Thierry}\
  \bibnamefont {Emonet}}, \bibinfo {author} {\bibfnamefont {Jose~MG}\
  \bibnamefont {Vilar}}, \bibinfo {author} {\bibfnamefont {Thomas~S}\
  \bibnamefont {Shimizu}}, \ and\ \bibinfo {author} {\bibfnamefont {Philippe}\
  \bibnamefont {Cluzel}},\ }\bibfield  {title} {\enquote {\bibinfo {title}
  {From molecular noise to behavioural variability in a single bacterium},}\
  }\href@noop {} {\bibfield  {journal} {\bibinfo  {journal} {Nature}\ }\textbf
  {\bibinfo {volume} {428}},\ \bibinfo {pages} {574--578} (\bibinfo {year}
  {2004})}\BibitemShut {NoStop}%
\bibitem [{\citenamefont {Korobkova}\ \emph {et~al.}(2006)\citenamefont
  {Korobkova}, \citenamefont {Emonet}, \citenamefont {Park},\ and\
  \citenamefont {Cluzel}}]{korobkova2006hidden}%
  \BibitemOpen
  \bibfield  {author} {\bibinfo {author} {\bibfnamefont {Ekaterina~A}\
  \bibnamefont {Korobkova}}, \bibinfo {author} {\bibfnamefont {Thierry}\
  \bibnamefont {Emonet}}, \bibinfo {author} {\bibfnamefont {Heungwon}\
  \bibnamefont {Park}}, \ and\ \bibinfo {author} {\bibfnamefont {Philippe}\
  \bibnamefont {Cluzel}},\ }\bibfield  {title} {\enquote {\bibinfo {title}
  {Hidden stochastic nature of a single bacterial motor},}\ }\href@noop {}
  {\bibfield  {journal} {\bibinfo  {journal} {Phys. Rev. Lett.}\ }\textbf
  {\bibinfo {volume} {96}},\ \bibinfo {pages} {058105} (\bibinfo {year}
  {2006})}\BibitemShut {NoStop}%
\bibitem [{\citenamefont {Wang}\ \emph {et~al.}(2017)\citenamefont {Wang},
  \citenamefont {Shi}, \citenamefont {He}, \citenamefont {Wang}, \citenamefont
  {Zhang},\ and\ \citenamefont {Yuan}}]{wang2017non}%
  \BibitemOpen
  \bibfield  {author} {\bibinfo {author} {\bibfnamefont {Fangbin}\ \bibnamefont
  {Wang}}, \bibinfo {author} {\bibfnamefont {Hui}\ \bibnamefont {Shi}},
  \bibinfo {author} {\bibfnamefont {Rui}\ \bibnamefont {He}}, \bibinfo {author}
  {\bibfnamefont {Renjie}\ \bibnamefont {Wang}}, \bibinfo {author}
  {\bibfnamefont {Rongjing}\ \bibnamefont {Zhang}}, \ and\ \bibinfo {author}
  {\bibfnamefont {Junhua}\ \bibnamefont {Yuan}},\ }\bibfield  {title} {\enquote
  {\bibinfo {title} {Non-equilibrium effect in the allosteric regulation of the
  bacterial flagellar switch},}\ }\href@noop {} {\bibfield  {journal} {\bibinfo
   {journal} {Nat. Phys.}\ }\textbf {\bibinfo {volume} {13}},\ \bibinfo {pages}
  {710--714} (\bibinfo {year} {2017})}\BibitemShut {NoStop}%
\bibitem [{\citenamefont {Figueroa-Morales}\ \emph {et~al.}(2020)\citenamefont
  {Figueroa-Morales}, \citenamefont {Soto}, \citenamefont {Junot},
  \citenamefont {Darnige}, \citenamefont {Douarche}, \citenamefont {Martinez},
  \citenamefont {Lindner},\ and\ \citenamefont {Cl{\'e}ment}}]{figueroa20203d}%
  \BibitemOpen
  \bibfield  {author} {\bibinfo {author} {\bibfnamefont {Nuris}\ \bibnamefont
  {Figueroa-Morales}}, \bibinfo {author} {\bibfnamefont {Rodrigo}\ \bibnamefont
  {Soto}}, \bibinfo {author} {\bibfnamefont {Gaspard}\ \bibnamefont {Junot}},
  \bibinfo {author} {\bibfnamefont {Thierry}\ \bibnamefont {Darnige}}, \bibinfo
  {author} {\bibfnamefont {Carine}\ \bibnamefont {Douarche}}, \bibinfo {author}
  {\bibfnamefont {Vincent~A}\ \bibnamefont {Martinez}}, \bibinfo {author}
  {\bibfnamefont {Anke}\ \bibnamefont {Lindner}}, \ and\ \bibinfo {author}
  {\bibfnamefont {Eric}\ \bibnamefont {Cl{\'e}ment}},\ }\bibfield  {title}
  {\enquote {\bibinfo {title} {3d spatial exploration by e. coli echoes motor
  temporal variability},}\ }\href@noop {} {\bibfield  {journal} {\bibinfo
  {journal} {Phys. Rev. X.}\ }\textbf {\bibinfo {volume} {10}},\ \bibinfo
  {pages} {021004} (\bibinfo {year} {2020})}\BibitemShut {NoStop}%
\bibitem [{\citenamefont {Ezhilan}\ and\ \citenamefont
  {Saintillan}(2015)}]{ezhilan2015transport}%
  \BibitemOpen
  \bibfield  {author} {\bibinfo {author} {\bibfnamefont {Barath}\ \bibnamefont
  {Ezhilan}}\ and\ \bibinfo {author} {\bibfnamefont {David}\ \bibnamefont
  {Saintillan}},\ }\bibfield  {title} {\enquote {\bibinfo {title} {Transport of
  a dilute active suspension in pressure-driven channel flow},}\ }\href@noop {}
  {\bibfield  {journal} {\bibinfo  {journal} {J. Fluid Mech.}\ }\textbf
  {\bibinfo {volume} {777}},\ \bibinfo {pages} {482--522} (\bibinfo {year}
  {2015})}\BibitemShut {NoStop}%
\bibitem [{\citenamefont {Frymier}\ \emph {et~al.}(1995)\citenamefont
  {Frymier}, \citenamefont {Ford}, \citenamefont {Berg},\ and\ \citenamefont
  {Cummings}}]{frymier1995three}%
  \BibitemOpen
  \bibfield  {author} {\bibinfo {author} {\bibfnamefont {Paul~D}\ \bibnamefont
  {Frymier}}, \bibinfo {author} {\bibfnamefont {Roseanne~M}\ \bibnamefont
  {Ford}}, \bibinfo {author} {\bibfnamefont {Howard~C}\ \bibnamefont {Berg}}, \
  and\ \bibinfo {author} {\bibfnamefont {Peter~T}\ \bibnamefont {Cummings}},\
  }\bibfield  {title} {\enquote {\bibinfo {title} {Three-dimensional tracking
  of motile bacteria near a solid planar surface.}}\ }\href@noop {} {\bibfield
  {journal} {\bibinfo  {journal} {PNAS}\ }\textbf {\bibinfo {volume} {92}},\
  \bibinfo {pages} {6195--6199} (\bibinfo {year} {1995})}\BibitemShut {NoStop}%
\bibitem [{\citenamefont {Singh}\ \emph {et~al.}(2021)\citenamefont {Singh},
  \citenamefont {Patteson}, \citenamefont {Maldonado}, \citenamefont
  {Purohit},\ and\ \citenamefont {Arratia}}]{singh2021bacterial}%
  \BibitemOpen
  \bibfield  {author} {\bibinfo {author} {\bibfnamefont {Jaspreet}\
  \bibnamefont {Singh}}, \bibinfo {author} {\bibfnamefont {Alison~E}\
  \bibnamefont {Patteson}}, \bibinfo {author} {\bibfnamefont {Bryan O~Torres}\
  \bibnamefont {Maldonado}}, \bibinfo {author} {\bibfnamefont {Prashant~K}\
  \bibnamefont {Purohit}}, \ and\ \bibinfo {author} {\bibfnamefont {Paulo~E}\
  \bibnamefont {Arratia}},\ }\bibfield  {title} {\enquote {\bibinfo {title}
  {Bacterial activity hinders particle sedimentation},}\ }\href@noop {}
  {\bibfield  {journal} {\bibinfo  {journal} {Soft Matter}\ }\textbf {\bibinfo
  {volume} {17}},\ \bibinfo {pages} {4151--4160} (\bibinfo {year}
  {2021})}\BibitemShut {NoStop}%
\bibitem [{\citenamefont {Maldonado}\ \emph {et~al.}(2023)\citenamefont
  {Maldonado}, \citenamefont {Pradeep}, \citenamefont {Ran}, \citenamefont
  {Jerolmack},\ and\ \citenamefont {Arratia}}]{maldonado2023sedimentation}%
  \BibitemOpen
  \bibfield  {author} {\bibinfo {author} {\bibfnamefont {Bryan O~Torres}\
  \bibnamefont {Maldonado}}, \bibinfo {author} {\bibfnamefont {Shravan}\
  \bibnamefont {Pradeep}}, \bibinfo {author} {\bibfnamefont {Ranjiangshang}\
  \bibnamefont {Ran}}, \bibinfo {author} {\bibfnamefont {Douglas}\ \bibnamefont
  {Jerolmack}}, \ and\ \bibinfo {author} {\bibfnamefont {Paulo~E}\ \bibnamefont
  {Arratia}},\ }\bibfield  {title} {\enquote {\bibinfo {title} {Sedimentation
  dynamics of passive particles in dilute bacterial suspensions: emergence of
  bioconvection},}\ }\href@noop {} {\bibfield  {journal} {\bibinfo  {journal}
  {arXiv preprint arXiv:2312.10226}\ } (\bibinfo {year} {2023})}\BibitemShut
  {NoStop}%
\bibitem [{\citenamefont {Massana-Cid}\ \emph {et~al.}(2024)\citenamefont
  {Massana-Cid}, \citenamefont {Maggi}, \citenamefont {Gnan}, \citenamefont
  {Frangipane},\ and\ \citenamefont {Di~Leonardo}}]{massana2024multiple}%
  \BibitemOpen
  \bibfield  {author} {\bibinfo {author} {\bibfnamefont {Helena}\ \bibnamefont
  {Massana-Cid}}, \bibinfo {author} {\bibfnamefont {Claudio}\ \bibnamefont
  {Maggi}}, \bibinfo {author} {\bibfnamefont {Nicoletta}\ \bibnamefont {Gnan}},
  \bibinfo {author} {\bibfnamefont {Giacomo}\ \bibnamefont {Frangipane}}, \
  and\ \bibinfo {author} {\bibfnamefont {Roberto}\ \bibnamefont
  {Di~Leonardo}},\ }\bibfield  {title} {\enquote {\bibinfo {title} {Multiple
  temperatures and melting of a colloidal active crystal},}\ }\href@noop {}
  {\bibfield  {journal} {\bibinfo  {journal} {arXiv preprint arXiv:2401.09911}\
  } (\bibinfo {year} {2024})}\BibitemShut {NoStop}%
\bibitem [{\citenamefont {Wysocki}\ \emph {et~al.}(2015)\citenamefont
  {Wysocki}, \citenamefont {Elgeti},\ and\ \citenamefont
  {Gompper}}]{wysocki2015giant}%
  \BibitemOpen
  \bibfield  {author} {\bibinfo {author} {\bibfnamefont {Adam}\ \bibnamefont
  {Wysocki}}, \bibinfo {author} {\bibfnamefont {Jens}\ \bibnamefont {Elgeti}},
  \ and\ \bibinfo {author} {\bibfnamefont {Gerhard}\ \bibnamefont {Gompper}},\
  }\bibfield  {title} {\enquote {\bibinfo {title} {Giant adsorption of
  microswimmers: duality of shape asymmetry and wall curvature},}\ }\href@noop
  {} {\bibfield  {journal} {\bibinfo  {journal} {Phys. Rev. E}\ }\textbf
  {\bibinfo {volume} {91}},\ \bibinfo {pages} {050302} (\bibinfo {year}
  {2015})}\BibitemShut {NoStop}%
\bibitem [{\citenamefont {Peng}\ \emph {et~al.}(2016)\citenamefont {Peng},
  \citenamefont {Lai}, \citenamefont {Tai}, \citenamefont {Zhang},
  \citenamefont {Xu},\ and\ \citenamefont {Cheng}}]{peng2016diffusion}%
  \BibitemOpen
  \bibfield  {author} {\bibinfo {author} {\bibfnamefont {Yi}~\bibnamefont
  {Peng}}, \bibinfo {author} {\bibfnamefont {Lipeng}\ \bibnamefont {Lai}},
  \bibinfo {author} {\bibfnamefont {Yi-Shu}\ \bibnamefont {Tai}}, \bibinfo
  {author} {\bibfnamefont {Kechun}\ \bibnamefont {Zhang}}, \bibinfo {author}
  {\bibfnamefont {Xinliang}\ \bibnamefont {Xu}}, \ and\ \bibinfo {author}
  {\bibfnamefont {Xiang}\ \bibnamefont {Cheng}},\ }\bibfield  {title} {\enquote
  {\bibinfo {title} {Diffusion of ellipsoids in bacterial suspensions},}\
  }\href@noop {} {\bibfield  {journal} {\bibinfo  {journal} {Phys. Rev. Lett.}\
  }\textbf {\bibinfo {volume} {116}},\ \bibinfo {pages} {068303} (\bibinfo
  {year} {2016})}\BibitemShut {NoStop}%
\bibitem [{\citenamefont {Baek}\ \emph {et~al.}(2018)\citenamefont {Baek},
  \citenamefont {Solon}, \citenamefont {Xu}, \citenamefont {Nikola},\ and\
  \citenamefont {Kafri}}]{baek2018generic}%
  \BibitemOpen
  \bibfield  {author} {\bibinfo {author} {\bibfnamefont {Yongjoo}\ \bibnamefont
  {Baek}}, \bibinfo {author} {\bibfnamefont {Alexandre~P}\ \bibnamefont
  {Solon}}, \bibinfo {author} {\bibfnamefont {Xinpeng}\ \bibnamefont {Xu}},
  \bibinfo {author} {\bibfnamefont {Nikolai}\ \bibnamefont {Nikola}}, \ and\
  \bibinfo {author} {\bibfnamefont {Yariv}\ \bibnamefont {Kafri}},\ }\bibfield
  {title} {\enquote {\bibinfo {title} {Generic long-range interactions between
  passive bodies in an active fluid},}\ }\href@noop {} {\bibfield  {journal}
  {\bibinfo  {journal} {Phys. Rev. Lett.}\ }\textbf {\bibinfo {volume} {120}},\
  \bibinfo {pages} {058002} (\bibinfo {year} {2018})}\BibitemShut {NoStop}%
\end{thebibliography}%

\end{document}